\documentclass[12pt]{article}
\usepackage{epsfig,picinpar,floatflt,amssymb}
\newcommand{\resection}[1]{\setcounter{equation}{0}\section{#1}}
\newcommand{\appsection}{\addtocounter{section}{1} \setcounter{equation}{0}
             \section*{Appendix \Alph{section}}}

\def\to{\rightarrow}
\def\goto{\longrightarrow}

\def\th{\theta}

\def\l{\lambda}

\def\vp{\varphi}

\def\bd{\begin{displaystyle}}
\def\ed{\end{displaystyle}}
\def\ba{\begin{array}}

\def\ea{\end{array}}
\def\EQ{\begin{equation}}
\def\EN{\end{equation}}
\def\bea{\begin{eqnarray}}
\def\eea{\end{eqnarray}}
\def\beano{\begin{eqnarray*}}
\def\eeano{\end{eqnarray*}}

\def\hs{\hspace{0.1in}}

\def\lab{\label}

\begin{document}
\oddsidemargin 5mm
\setcounter{page}{0}
\newpage     
\setcounter{page}{0}
\begin{titlepage}
\begin{flushright}
ISAS/EP/97/106 \\
IC/97/129
\end{flushright}
\vspace{0.5cm}
\begin{center}
{\large {\bf Non-Integrable Aspects of the Multi-frequency \\
Sine-Gordon Model}}\footnote{Work done under partial support of 
the EC TMR Programme {\em Integrability, non--perturbative effects 
and symmetry in Quantum Field Theories}, grant FMRX-CT96-0012} \\
\vspace{1.5cm}
{\bf G. Delfino$^{a}$ and G. Mussardo$^{b,c,d}$} \\
\vspace{0.8cm}
$^a${\em Laboratoire de Physique Th\'eorique,
Universit\'e de Montepellier II \\
Pl. E. Bataillon, 34095 Montpellier, France} \\
$^b${\em International School for Advanced Studies, Via Beirut 2-4, 
34013 Trieste, Italy} \\ 
$^c${\em Istituto Nazionale di Fisica Nucleare, Sezione di Trieste}\\
$^d${\em International Centre of Theoretical Physics \\
Strada Costiera 12, 34014 Trieste }\\
\end{center}
\vspace{6mm}
\begin{abstract}
\noindent
We consider the two-dimensional quantum field theory of a scalar field  
self-interacting via two periodic terms of frequencies $\alpha$ and $\beta$.
Looking at the theory as a perturbed Sine-Gordon model, we use Form Factor 
Perturbation Theory to analyse the evolution of the spectrum of particle 
excitations. We show how, within this formalism, the non-locality of the 
perturbation with respect to the solitons is responsible for their confinement 
in the perturbed theory. The effects of the frequency ratio $\alpha/\beta$ 
being a rational or irrational number and the occurrence of massless flows 
from the gaussian to the Ising fixed point are also discussed. 
A generalisation of the Ashkin-Teller model and the massive Schwinger model 
are presented as examples of application of the formalism.
\end{abstract}
\vspace{5mm}
\end{titlepage}
\newpage

\setcounter{footnote}{0}
\renewcommand{\thefootnote}{\arabic{footnote}}

\resection{Introduction}

\,

For its attractive features, both at classical and quantum level, the 
$(1+1)$-dimensional Sine-Gordon (SG) model has always played a 
prominent role in the investigation of theoretical non-perturbative 
aspects of quantum field theory (QFT) [1---7] as well as in the 
concrete discussion of innumerable physical systems: equations of
Sine-Gordon type are in fact relevant in quasi one-dimensional 
charge density waves of strongly correlated electrons and in 
statistical mechanics aspects of one--dimensional quantum chains 
\cite{Luther}, in self-induced trasparency effects of non--linear 
optics \cite{optics,Leclair} or spin wave propagation in the quantum 
liquid phase of $^3$He \cite{Leggett}, just to mention a few.  
The lagrangian of the model may be written as 
\EQ
{\cal L}_{SG} \,=\,\frac{1}{2} (\partial_{\mu} \vp)^2 + 
\mu \,\cos(\beta \vp) 
\,\,\,. 
\label{sg}
\EN 

In many of the physical applications, the mapping onto the pure SG model 
is, however, only an approximation and therefore is expected to hold only 
in some special regions of the physical parameter space. A more complete
and refined description of several physical effects often requires that 
a lagrangian with two or more periodic interaction terms is considered. A  
simple example is provided by a system of two Ising models coupled 
both by thermal and magnetic operators, system which may be considered 
as a generalization of the Ashkin-Teller model\footnote{This model will be
discussed in some details in Section 5 of this paper.}. Other examples,
arising from different physical contexts, are given by the aforementioned
cases of spin waves propagating in anisotropic magnetic liquids or ultra-short 
optical pulses propagating in resonant degenerate medium \cite{Bullough}. 
The purpose of this paper is to analyse the quantum field theory aspects 
of this kind of generalisation of the SG model and in particular the case 
in which two periodic interactions are present, i.e. to study the QFT 
defined by the lagrangian\footnote{Meaning and restrictions 
on the parameters entering eq.\,(\ref{lagrangian}) will be the subject of
Section 2.} 
\EQ
{\cal L} \,= \,
\frac{1}{2}(\partial_\mu\varphi)^2 + \mu\cos\beta\varphi +
\lambda\cos(\alpha\varphi+\delta) \,\,\,.
\label{lagrangian}
\EN
Apart from the practical motivations we have just mentioned, there are also
some theoretical issues which make the analysis of (\ref{lagrangian}) an 
interesting problem. Let us try to describe them.

The addition of a new interaction term to the lagrangian (\ref{sg}) spoils 
one of the crucial properties of the SG model, i.e. its integrability. 
As a consequence, the resulting two-frequency model (\ref{lagrangian}) is 
not exactly solvable and perturbation theory is basically the only
available tool for a quantitative study. Of all the different perturbative
approaches that could be applied to the problem at hand, the one 
where the new periodic term is viewed as perturbing a pure SG model
is probably the least conventional but the most
convenient.  Perturbation of integrable models was discussed in
Ref.\,\cite{DMS}: its theory relies on the fact that the matrix
elements of local operators between asymptotic states of the unperturbed 
(integrable) theory are exactly computable. The corrections to the masses 
of the particles of the unperturbed theory and to the scattering
amplitudes can be computed in terms of the matrix elements of the perturbing 
operator in much the same way as in quantum mechanics. Since the
matrix elements of the local operators are usually known as Form 
Factors, we refer to this perturbative approach as Form Factor 
Perturbation Theory (FFPT). The Form Factors of the SG model were 
computed in Refs.\,\cite{KW,Smirnov,Luky} and consequently the technical 
information required for these perturbative calculations is available. 
However, our interest is not to perform specific perturbative computations,
but rather to illustrate some general features of the perturbed model 
and the use of the FFPT to describe it. 

We are particularly interested in the spectrum of the physical
excitations. The invariance of the Lagrangian (\ref{sg}) under 
the discrete transformation $\varphi \goto \varphi + 2 \pi q/\beta$, 
$q \in \mathbb{Z}$ implies that the elementary excitations of the pure SG
model are solitons interpolationg between adjacent degenerate minima of the
potential. Extra interactions will generally remove the degeneracy among the 
different classical vacua and therefore these topological excitations 
are expected to disappear from the spectrum of asymptotic states of 
the perturbed theory. Therefore it is essential to ascertain 
whether the FFPT accounts for such a dramatic evolution in the particle
spectrum of the theory. As we will show, the answer is indeed affirmative and
the phenomenon of confinement of the original topological 
excitations emerges in the FFPT as a consequence of the mutual non-locality
between the perturbing operator and the field which interpolates the 
confined particle. Notably, the non--local mechanism responsible for the 
confinement of the topological excitations proves to be a general 
feature of many two-dimensional theories: it causes, for instance, 
the disappearence of the kinks from the spectrum of the low-temperature 
phase of the Ising model as soon as an external magnetic field is turned on
\cite{DMS,Mccoy-Wu}.

The role of the quantum vacuum of the theory is particularly relevant 
for the analysis below. In the pure SG model, the quantum vacuum is
selected from the infinite number of classical equivalent vacua 
through the spontaneous symmetry breaking mechanism. At the
operational level, this involves selecting arbitrarily one of the
classical minima as the quantum vacuum and constructing the space of 
asymptotic states as excitations above it. Physical quantities such as 
scattering amplitudes, Form Factors, etc. are then computed as matrix 
elements on this space of states. In the perturbed theory, the degeneracy 
of the classical vacua is explicitely broken and the quantum vacuum is
uniquely defined. A fundamental requirement for the application of the 
FFPT is to ensure that, when the perturbation is adiabatically switched
off, the vacuum matches the one previously chosen for the 
computations in the unperturbed theory. This ``adiabatic condition'' 
corresponds to the general requirement that the vacuum of a quantum 
field theory in the presence of a spontaneously broken symmetry must be
identified by taking the limit in which a symmetry breaking perturbation 
is made to vanish.

The vacuum of the perturbed theory is however unique and 
the above considerations apply, only if the periods of the two 
interaction terms are incommensurate. In the case of rational 
frequency ratio, the survival of a subset of degenerate classical 
vacua in the perturbed potential must be given appropriate 
consideration. It is already clear from this observation that the 
theory will exhibit an extremely irregular behaviour as a function 
of the frequency ratio.

In the non--integrable model (\ref{lagrangian}), the two periodic
interactions clearly play a symmetric role and each of the cosine 
interactions can be viewed as a deformation of the integrable theory 
defined by the other: as discussed in more detail in Section 3, there 
is a dimensionless variable  
\EQ
\eta \equiv \lambda\mu^{-(1-\Delta_\alpha)/(1-\Delta_\beta)} =
\lambda\mu^{-(8\pi-\alpha^2)/(8\pi-\beta^2)}\,\, ,
\label{rg}
\EN
which characterises the two perturbative regimes of the model
(\ref{lagrangian}). The first regime is obtained in the limit 
$\eta \rightarrow 0$, while the other is reached for $\eta 
\rightarrow \infty$, the latter by just swapping the role played 
by the two operators. The possibility to control the changes of the 
spectrum in both perturbative limits allows one to deduce 
interesting information about its evolution in the intermediate,
non-perturbative region. In certain cases, there may be a topological 
excitation in one limit which is no longer present in the other. 
When this is the case, the very nature of topological excitations requires 
that a change in the vacuum structure of the theory takes place 
somewhere in the non-perturbative region, namely that a phase 
transition occurs\footnote{This phenomenon was discussed in 
these terms long ago by S. Coleman in his classic study of the massive 
Schwinger model \cite{Coleman2}.}. Lines of phase transition are then 
expected to appear in the multi--frequency SG model for particular
values of the parameters: they generally correspond to Renormalisation 
Group (RG) trajectories along which the system flows from the gaussian 
fixed point with central charge $C=1$ to the Ising fixed point with 
central charge $C=1/2$.

The paper is organised as follows. The next section reviews several 
results of the pure SG model which will be used later. The two-frequency 
model is introduced in Section\,3: some restrictions on the parameters 
of the theory dictated by conformal perturbation theory and the 
aforementioned adiabatic condition will be discussed. The Form Factor 
Perturbation Theory, the relation between non-locality and 
soliton confinement, the evolution of the particle spectrum and the 
occurrence of phase transition are analysed in Section\,4. The
system of two--layer Ising model, alias the generalized Ashkin-Teller 
model, and the massive Schwinger model are presented as examples of 
specific applications of the formalism in Section\,5. Summary of the 
results and conclusions are in Section\,6. The paper has 
also two appendices: the first contains a derivation of the 
soliton-antisoliton form factors of exponential operators at the 
reflectionless points of SG model while the second presents an 
analysis of soliton confinement within the semiclassical approximation. 

\resection{Few Notions of the Sine--Gordon Model}

In this section we briefly review some basic features of the
pure SG model and establish several notations which will be used
in the following. The SG model is a relativistic theory in $(1 + 1)$ 
dimensions for a scalar bosonic field $\vp(x)$ with the lagrangian 
given in eq.\,(\ref{sg}). At the classical level, the model possesses
an infinite degeneracy of the vacuum states which can be labelled by 
a relative integer $q$: finite energy configurations must necessarely 
approach one of the vacuum values $\frac{2 \pi}{\beta} q_-$ for 
$x \to - \infty$ and another one $\frac{2 \pi}{\beta} q_+$ for 
$x \to + \infty$. The difference $Q \equiv q_+ - q_-$ plays a role 
of topological charge which characterises inequivalent sectors of the 
theory. As well known, the model is completly solvable both at the 
classical and the quantum level. At the classical level, all solutions 
of the equation of motion can be computed by means of the inverse 
scattering method, as reviewed for instance in \cite{IEEE}. Building 
blocks for expressing all the classical solutions are the static soliton 
and antisoliton configurations with $Q = \pm 1$: their explicit expressions 
are given by 
\EQ
\vp_{s,\overline s} (x) \,=\,\frac{4}{\beta} \arctan 
[\exp(\pm\beta\sqrt{\mu}\,x)] \,\, , 
\label{soliton}
\EN 
where $+$ refers to the soliton and $-$ to the antisoliton.  
Multi--solitons are then obtained by the non--linear superposition 
principle provided by the B\"{a}cklund transformations. For the 
purposes of this paper, it may be convenient to think of soliton 
configurations as step functions with jumps $\pm 2\pi/\beta$ 
(Figure 1). 

At the quantum level, a normal ordering of the cosine interaction term in 
the lagrangian (\ref{sg}) is sufficient to get rid of the only diverging 
diagrams (tadpoles) in the theory. From the conformal perturbation 
theory viewpoint, the SG model can be considered as a deformation of 
the gaussian fixed point action 
\EQ
{\cal A}_{Gaussian} \,=\,\int d^2x \,\frac{1}{2} \left(\partial_{\nu} 
\varphi \right)^2 \,\,\, ,
\label{gaussian}
\EN 
(with central charge $C=1$) by the operator $\cos\beta\varphi$ with 
conformal dimension $\Delta_\beta = \beta^2/8\pi$. The latter is relevant 
for $\beta^2 < 8 \pi$, and in this range the theory presents a massive 
phase\footnote{The limitation $\beta^2 < 8 \pi$ will be 
always assumed to hold in the following.}. The exact particle spectrum 
of the theory \cite{DHN,ZZ} consists of a pair of soliton and antisoliton 
excitations with equal mass $M$ and a number $N =
\left[\frac{\pi}{\xi}\right]$ of soliton--antisoliton bound states 
$B_n$ (breathers) with masses 
\EQ
m_k \,=\, 2 M \sin\left(\frac{k \xi}{2}\right) \,\,\, , 
\,\,\,\,\, k =1,2,\ldots,< \frac{\pi}{\xi} \,\,\, ,
\label{masses}
\EN
where 
\EQ
\xi \,=\, \frac{\pi\beta^2}{8\pi - \beta^2} \,\,\,.
\label{rencoupling}
\EN 
The elementary field $\vp$ entering the lagrangian (\ref{sg}) may be regarded 
as associated to the lightest breather $B_1$. Note, however, that the
$k$-th bound state crosses the soliton--antisoliton threshold when 
$\xi= \pi/k$, and for $\xi \geq \pi/k$ it leaves the physical spectrum. 
Hence, for $\xi \geq \pi$, the spectrum consists of soliton and antisoliton 
only and does not contain any particle associated to the field $\vp$.

An additional fundamental result about SG model is its exact equivalence 
with the massive Thirring model \cite{Coleman},
namely with the theory of a self-interacting Dirac fermion defined by the 
Lagrangian
\EQ
{\cal L}_{MTM}=\bar{\psi}(i\gamma^\mu\partial_\mu-m)\psi-
\frac{g}{2}(\bar{\psi}\gamma^\mu\psi)^2\,\,.
\label{mtm}
\EN
The correspondence between the two models is specified by the bosonisation 
rules and by the following relation among the couplings of the two models 
\EQ
g\,=\,\pi\frac{4\pi-\beta^2}{\beta^2}\,\,\,.
\label{g}
\EN
The exact $S$--matrix relative to factorizable and elastic scattering
processes of the SG model has been determined in \cite{ZZ}. We refer 
the reader to this paper for all details about this subject. The matrix
elements of local operators between the asymptotic states of the SG model 
have also been computed by using the so called Form Factor bootstrap 
approach \cite{KW,Smirnov}. Recently, Lukyanov used a technique based on free 
field representation to compute the Form Factors of the exponential
operators $e^{i\alpha\varphi}$ \cite{Luky}. Since these Form Factors are of
particular interest in the present paper, we show in Appendix A how they 
can be easily computed in the ordinary bootstrap framework, at least for 
the special values of the coupling $\xi=\pi/k$ where the $S$-matrix is 
reflectionless and the technical problem gets considerably simplified.

\resection{The Two-frequency Sine-Gordon Model}

In this section we will derive and illustrate some limitations on the
real parameters $\alpha$, $\beta$ and $\delta$ entering the quantum action 
\EQ
{\cal A} \,= \,\int\,d^2 x
\left[ \frac{1}{2}(\partial_\mu\varphi)^2 + 
\mu\cos\beta\varphi+
\lambda\cos(\alpha\varphi+\delta) \right]\,\,\,. 
\label{actiondsg}
\EN
We can always assume that $\alpha$, $\beta$, $\mu$ and
$\lambda$ are positive, since this situation can be arranged by 
using appropriate redefinitions of $\varphi$ and $\delta$. 

There are at least two ways of viewing the above theory, both of 
which provide some useful insight to its physical content. The first 
regards (\ref{actiondsg}) as a perturbation of the gaussian action 
(\ref{gaussian}) by means of the two scaling operators ${\cal O}_1 
= \cos\beta\varphi$ and ${\cal O}_2 = \cos(\alpha\varphi+\delta)$, 
with conformal dimensions $\Delta_\beta = \beta^2/8\pi$ and 
$\Delta_\alpha = \alpha^2/8\pi$, respectively. The coupling constants $\mu$
and $\lambda$ have scaling dimensions $(mass)^{2(1-\Delta_\beta)}$ 
and $(mass)^{2(1-\Delta_\alpha)}$, respectively and therefore 
the parameter $\eta$ which was defined in (\ref{rg}) forms a dimensionless
combination which labels the different RG flows originating from the
gaussian fixed point. The relevant nature of both operators then implies
\EQ
\begin{array}{c}
\alpha^2 < 8 \pi \,\,\,; \\
\beta^2 < 8\pi \,\,\,.
\end{array}
\label{1request}
\EN
An additional constraint on the two parameters $\alpha$ and $\beta$ 
derives from the renormalizability of the action (\ref{actiondsg}), 
an aspect which can be efficiently studied by conformal perturbation 
theory \cite{Aliosha}. To illustrate this point, let us briefly 
discuss the pattern of renormalisation of a scaling operator $\Phi(x)$ 
in a theory obtained by perturbing a conformal action\footnote{We 
consider for simplicity the case of perturbation by a single scaling 
operator, but the following considerations are easily extended to more 
general situations.} by a relevant operator $\phi$ of conformal dimension 
$\Delta$ 
\EQ
{\cal A}_g \,=\, {\cal A}_{CFT} + g\int d^2x\,\phi(x) \,\,\,.
\EN
Let $X$ denote a generic product of operators and consider the 
usual perturbative expansion of the correlator
\EQ
\langle X\,\Phi(0)\rangle \,=\, \langle X\,\Phi(0)\rangle_{CFT} +
g \int_{\epsilon<|x|<R}d^2x\langle X\,\Phi(0)\,\phi(x)\rangle_{CFT}
+ O(g^2)\,\,\,.
\label{cpt}
\EN
$\epsilon$ and $R$ are ultraviolet and infrared cutoffs, respectively, and 
the correlators on the right hand side of the above equation are computed 
at the conformal point. The integral in Eq.\,(\ref{cpt}) is ultraviolet 
divergent only if the conformal OPE  
\EQ
\phi(x)\,\Phi(0) \,=\, \sum_k C_{\phi\Phi}^k 
|x|^{2(\Delta_k - \Delta_\Phi - \Delta)}\, A_k(0)
\EN
contains operators $A_k$ with conformal dimension $\Delta_k$ such that
\EQ
\gamma_k \equiv \Delta_k - \Delta_\Phi - \Delta+1\leq 0\,\,\,.
\label{gamma}
\EN
If this is the case, a first order UV finite correlator $\langle
X\,\Phi'(0)  \rangle$ is obtained by defining the renormalised operator
\EQ
\Phi' \,\equiv \, \Phi + g \sum_k b_k\epsilon^{2\gamma_k}A_k + O(g^2)\,\,,
\EN
where $b_k = -\pi C_{\phi\Phi}^k/\gamma_k$ and the sum runs only over the 
operators $A_k$ satisfying the condition (\ref{gamma}). The last equation 
shows that the renormalisation of UV divergences induces a mixing of the 
operator $\Phi(x)$ with a finite number of operators having smaller
conformal dimensions.

The above general considerations are easily applicable to
perturbations of the gaussian action, with the spectrum of primary 
operators spanned by the exponentials $e^{i\alpha\varphi(x)}$ of  
conformal dimension $\Delta_\alpha = \alpha^2/8\pi$ and their OPE 
given by 
\EQ
e^{i\alpha_1\varphi(x)}\,e^{i\alpha_2\varphi(0)}
\,=\, |x|^{\alpha_1\alpha_2/2\pi}\, e^{i(\alpha_1+\alpha_2)\varphi(0)}\,\,.
\EN
Using eq.\,(\ref{gamma}), it is easy to see that the two perturbing terms 
in (\ref{actiondsg}) do not mix with other operators (so that no extra
terms are generated in the action (\ref{actiondsg})) if the parameters 
$\alpha$ and $\beta$ satisfy the inequality  
\EQ
\alpha \beta < 4\pi \,\,\,.
\label{2request}
\EN
If the above condition is not fulfilled, the action (\ref{actiondsg}) 
results unstable under renormalization, i.e. additional counterterms 
$e^{\pm i(\alpha-\beta)\varphi}$ should already be included at the
first  order in $g$ so as to have a consistent theory\footnote{In the
particular case $\alpha = \beta$ this simply leads to the introduction of an
inessential additive constant in the action (\ref{actiondsg}).}.

Let us now consider the second approach to the QFT (\ref{actiondsg}) --
the one we will mostly adopt in this paper -- which consists of viewing it
as a deformation of a pure Sine-Gordon action. Although the two interaction
terms play a completely symmetric role, in the following we will assume
$\lambda\cos(\alpha \varphi + \delta)$ to be the perturbing term and derive
some restrictions  on the parameter $\delta$. 

When $\lambda = 0$, the pure SG potential has a periodicity $2 \pi/\beta$
and therefore exhibits an infinite number of classical degenerate vacua 
placed at $\varphi = 2 \pi q/\beta $. At the quantum level, however,
spontaneous symmetry breaking selects a unique quantum vacuum, which 
by convention we choose to be at the origin, $\varphi =0$. Let us now 
switch the perturbation on. Our analysis begins by considering 
the case when the frequency ratio 
\EQ
\omega \equiv \frac{\alpha}{\beta} 
\label{frequencyratio}
\EN 
is a rational number, $m/n$, with $m$ and $n$ coprime positive integers. 
In this case, the new potential acquires a periodicity $2 \pi
n/\beta$. For $\lambda$ very small, the perturbed potential presents, within 
the first period, $n$ relative minima at $\varphi_k = 2 \pi k/\beta + O(\l)$,
$k=0,1,\ldots, n-1$ (Figure 2). Let the absolute minimum of the perturbed 
potential be located nearby $\varphi_{\tilde k} = 2 \pi \tilde k/\beta$.
According to the adiabatic condition, we have then to shift 
\EQ
\vp(x) \goto \vp - 2 \pi \tilde k/\beta \,\, , 
\label{shift}
\EN 
in order to move the vacuum back near to the origin: the action
(\ref{actiondsg}) is intended to precisely address to this situation. 
Hence, the only values of $\delta$ which lead to inequivalent quantum 
field theories -- those which do not differ by a simple shift in the 
location of the vacuum -- are those given by 
\EQ
|\delta|\,\leq\,\frac{\pi}{n} \,\,\,. 
\label{3request}
\EN
These are in fact the values for which the absolute minimum of the 
perturbed potential remains located close to $\varphi = 0$. At the
perturbative level in $\l$, eq.\,(\ref{3request}) is simply derived 
from the requirement that the variation of the potential at the minimum 
close to $\varphi = 0$ is larger (in absolute value) than at the 
other $(n - 1)$ minima. When the bound (\ref{3request}) is saturated, 
one of the $(n-1)$ minima and the one close to the origin degenerate 
(Figure 3).

Let us now consider the case in which the ratio $\omega$ of the two
frequencies is an irrational number. Since any irrational 
number can be approximated with arbitrary precision by the sequence of 
rational approximants provided by its continued fraction expansion, this 
case may be viewed as a particular limit of the previous situation. Let  
\EQ
\frac{m_i}{n_i} \equiv  
\frac{1}{r_1 + \frac{1}{r_2 + \frac{\cdots}{r_{i-1}+ \frac{1}{r_i}}}}
\label{fraction}
\EN
be the $i$-th approximation of the irrational number $\omega$. 
Both $m_i$ and $n_i$ become larger by increasing the index 
$i$ of the sequence\footnote{A well known example of this statement is
provided by the sequence $F_{i}/F_{i+1}$ of the Fibonacci numbers which
converges  to the golden ratio $(\sqrt{5}-1)/2$.}. Since the restriction
(\ref{3request}) only involves the denominator $n_i$ of the rational 
number, the range of values of $\delta$ which give rise to inequivalent 
theories shrinks consequently to zero in the irrational case. 

In conclusion, the family of RG trajectories in the $\lambda \mu$--plane 
which are labelled by the dimensionless parameter $\eta$ acquires, 
somehow, a ``fractal'' nature as a function of the frequency ratio
$\omega$: while no further parametric dependence is present for 
irrational values of $\omega$, on the contrary a multiplicity of 
trajectories -- labelled by the inequivalent values of $\delta$
(\ref{3request}) -- results for each rational value of the frequency 
ratio. 

The varied behaviour of the theory (\ref{actiondsg}) -- depending on  
whether $\omega$ is a rational or an irrational number -- can be 
understood in a number of ways. In the presence of an irrational
frequency ratio, the perturbed potential has a very irregular shape: 
it is no longer periodic and possesses an infinite number of 
inequivalent relative minima. The values of the potential at these minima
form a continuous spectrum above the absolute minimum value
$V_{\makebox{\rm min}} = -\mu-\lambda$. Suppose that the absolute minimum
lies at $\varphi = \varphi'$ for a given value of $\delta$ (e.g. $\varphi'
= 0$ for $\delta=0$). An infinitesimal variation of the parameter 
$\delta$ induces a corresponding variation of the potential at $\varphi'$.
Since the values of the potential at the minima form a continuous 
spectrum, the absolute minimum will no longer be located near $\varphi'$
but will jump to another point $\varphi''$ which is generally far 
from $\varphi'$. Hence, a variation of $\delta$ simply amounts to a shift
of the vacuum from $\varphi'$ to $\varphi''$. This in turn means 
that $\delta$ is an irrelevant parameter in the case of irrational
frequency ratio and can  be safely set to zero. 

The above considerations can also be rephrased in more geometrical
terms, as follows. Let us introduce an additional field $\zeta(x)$ and 
write the potential of the lagrangian (\ref{lagrangian}) as 
\EQ
V(\varphi,\zeta) \,=\, - 
\mu \cos\beta\varphi - 
\lambda \cos(\alpha\zeta + \delta) \,\,\,. 
\EN
The original theory is of course recovered by the identification 
\EQ
\zeta \equiv \varphi \,\,\,.
\label{identification}
\EN 
Independently from the nature of the ratio $\omega$, the potential
$V(\varphi,\zeta)$ presents a double periodicity $\left(\frac{2 \pi}
{\beta},\frac{2\pi}{\alpha}\right)$ as a function of the two variables 
$\varphi$ and $\zeta$. Hence, this function is naturally defined on a 
compact torus ${\cal T} = \left\{(\varphi,\zeta): -\frac{\pi}{\beta} 
\leq \varphi \leq  \frac{\pi}{\beta}\,;\,-\frac{\pi}{\alpha} \leq \zeta
\leq \frac{\pi}{\alpha}\right\}$, and in this domain has an absolute 
minimum located at  $P^\star =  (0, -\frac{\delta}{\alpha})$. 
To this point, the rational or irrational nature of $\omega$ has played no
role. Only by enforcing the identification (\ref{identification}) does 
the difference between the two cases emerge. Note in fact that on the
torus ${\cal T}$ eq.(\ref{identification}) defines a family of straight
lines which all have slope $1$ (Figure 4). These straight lines may be 
viewed as the motion of a geometrical point bouncing on the boundaries of
the torus and reappearing on the opposite sides via its periodicity. When
$\omega$ is a rational number $m/n$, the motion of this point repeats 
over along the finite number $n+m-1$ of these diagonal lines (Figure
4.a). The motion may pass through the minimum $P^\star$ or remain at 
a finite distance from it, which depends on the value of $\delta$. It is
however clear that in the rational case all values $|\delta|<\frac{\pi}{n}$ 
are inequivalent and that when $\delta$ 
exceeds this bound we can go back to the case already considered
by simply redefining the fundamental domain ${\cal T}$ of the
torus. In the irrational case, on the contrary, the motion of the 
geometrical point is ergodic and therefore spans the entire torus 
${\cal T}$ (Figure 4.b). Hence, it passes arbitrarly close to
the minimum $P^\star$, {\em regardless } on the value of
$\delta$.  Consequently, the theory with $\delta \neq 0$ is physically 
indistinguishable from the one with $\delta = 0$, the latter value 
only reflects the choice of $\varphi = 0$ as the vacuum state. 

\resection{Non-locality, Soliton Confinement and Phase Transitions}

In this section we will consider the quantum theory (\ref{actiondsg}) 
as a perturbed Sine-Gordon model with the purpose of investigating some 
features of the evolution of the particle spectrum of the theory as 
a function of the parameter $\eta$ defined in (\ref{rg}). 

\subsection{Form Factor Perturbation Theory and Non--locality}

Perturbation theory around integrable models was discussed in
Ref.\,\cite{DMS}, where it was shown how the corrections to the mass 
spectrum and scattering amplitudes can be expressed in terms of the 
exactly computable Form Factors of the perturbing operator on the 
asymptotic states of the unperturbed theory. For the purposes of this paper 
we only need to point out some of their properties and few basic 
results of Ref.\,\cite{DMS}. 

Let $\Psi(x)$ be the scalar operator which perturbs an integrable
action, 
\EQ
{\cal A} \,=\,{\cal A}_{\rm int} + \lambda \int\,d^2x\,\Psi(x) \,\,\,.
\label{non}
\EN 
One of the first effects of moving away from integrability is a
change in the spectrum of the theory: the first order correction to the
mass of a particle $a$ belonging to the spectrum of the unperturbed theory
is given by 
\EQ
\delta m_a^2\simeq 2\lambda F_{a\bar{a}}^\Psi(i\pi)\,\,,
\label{deltam}
\EN
where the particle-antiparticle Form Factor of the operator $\Psi(x)$,
defined as the matrix element\footnote{The standard parameterisation of 
two-dimensional on mass-shell momenta in terms of rapidities is adopted:
$p_a^0=m_a\cosh\th_1$, $p_a^1=m_a\sinh\th_1$, $\th\equiv\th_1-\th_2$.}
\EQ
F_{a\bar{a}}^\Psi(\th)\equiv\langle 0|\Psi(0)|
a(\th_1)\bar{a}(\th_2)\rangle\,\, ,
\label{aabar}
\EN
is introduced. 

In integrable theories Form Factors of a generic scalar operator
${\cal O}(x)$ satisfy manageable functional equations in virtue 
of the simple form assumed by the unitarity and crossing symmetry 
equations \cite{KW,Smirnov}. For the two--particle case, we have 
\EQ 
F_{a\bar{a}}^{\cal O}(\th)=S_{a\bar{a}}^{b\bar{b}}(\th)
F_{\bar{b}b}^{\cal O}(-\th)\,\,,
\label{uni}
\EN
\EQ
F_{a\bar{a}}^{\cal O}(\th+2i\pi)=e^{-2i\pi\gamma_{{\cal O},a}}
F_{\bar{a}a}^{\cal O}(-\th)\,\,.
\lab{cross}
\EN

The first of these equations expresses the fact that in an integrable 
theory the two-particle threshold is the only unitarity branch point in 
the plane of the Mandelstam variable $s=(p_a+p_{\bar{a}})^2 = 4m_a^2 
\cosh^2(\th/2)$, the discontinuity across the cut being determined 
by the two-body scattering amplitude $S_{a\bar{a}}^{b\bar{b}}$. 

In the second equation the explicit phase factor $e^{-2i\pi \gamma_{
{\cal O},a}}$ is inserted to take into account a possible semi-locality 
of the operator which interpolates the particle $a$ (i.e. any operator
$\varphi_a$ such that $\langle 0|\varphi_a|a\rangle\neq 0$) with respect to
the operator ${\cal O}(x)$ \footnote{Consistency of eq.\,(\ref{cross}) 
requires $\gamma_{{\cal O},\bar{a}} = - \gamma_{{\cal O},a}$.}.
We must bear in mind that two operators ${\cal O}_1$ and ${\cal O}_2$
are said to be mutually non-local if the euclidean correlator 
$\langle\ldots{\cal O}_1(x_1){\cal O}_2(x_2)\ldots\rangle$ is not a 
single valued function of $x_1$ and $x_2$. In the particular case in
which the correlator simply acquires a phase factor under the analytic
continuation bringing $x_2$ to its original position along a path
encircling $x_1$, then the operators ${\cal O}_1$ and ${\cal O}_2$ 
are said to be semi-local with respect to each other. Equation
(\ref{cross}) can be regarded as a momentum space version of this 
statement for the operators ${\cal O}(x)$ and $\varphi_a$(x). When
$\gamma_{{\cal O},a}  =0$, there is no crossing symmetric counterpart 
to the unitarity cut but when $\gamma_{{\cal O},a}\neq 0$, there is 
a non-locality discontinuity in the $s$-plane with $s=0$ as branch 
point. In the rapidity parameterisation there is however no cut because the 
different Riemann sheets of the $s$-plane are mapped onto different 
sections of the $\th$-plane; the branch point $s=0$ is mapped onto 
the points $\th=\pm i\pi$ which become the locations of simple 
{\em annihilation} poles, whose residues are given by 
\cite{Smirnov} (see also \cite{Yurov})  
\EQ
-i\,{\mbox Res}_{\th=\pm i\pi}F_{a\bar{a}}^{\cal O}(\th)=
(1-e^{\mp 2i\pi\gamma_{{\cal O},a}})\langle 0|{\cal O}|0\rangle\,\,.
\label{pole}
\EN

With the above information, let us now proceed to the perturbative
analysis of the action (\ref{actiondsg}). As before, in the following we will 
refer to the case $\eta \ll 1$, so that the role of the perturbing operator
$\Psi$ in (\ref{deltam}) is played by $\cos(\alpha\varphi+\delta)$. It is
understood that a similar analysis can be repeated in the opposite limit
$\eta \goto \infty$, with $\mu\cos\beta\varphi$ assumed as the perturbation.

\subsection{Fate of the solitons}

The particle spectrum of the unperturbed theory at $\eta=0$ consists of 
a soliton-antisoliton pair and -- for $\beta^2<4\pi$ -- a certain number of 
breathers. 

Breathers are mutually local with respect to the elementary Sine-Gordon 
field $\varphi$, and consequently to the perturbing
operator $\cos(\alpha\varphi+\delta)$. Thus, formula (\ref{deltam}) can be
safely used to explicitely compute the first order correction to the 
unperturbed masses (\ref{masses})\footnote{The Form Factors which are
actually needed can be found in \cite{Luky,KM}.}. 

The situation is quite different for the solitonic sector of
the Hilbert space. The soliton is an elementary excitation of the
unperturbed theory which interpolates between two constant configurations 
of the field $\varphi$ differing by $2\pi/\beta$. This means that $\varphi 
\goto \varphi + 2\pi/\beta$ across a soliton configuration. As a consequence, 
the exponential operator $e^{i\alpha\varphi}$ is semi-local with respect 
to the soliton with semi-locality index $\gamma_{\alpha,s}=\alpha/\beta$. 
According to equation (\ref{pole}), the soliton-antisoliton Form Factor 
of the perturbing operator $\Psi = \cos(\alpha\varphi + \delta)$ 
contains annihilation poles in $\th=\pm i\pi$ whose residues are 
immediately computed as\footnote{This result refers to the choice 
$\langle 0|\varphi|0\rangle = 0$ for the unperturbed theory. 
In this case the unperturbed theory is invariant under the 
reflection $\varphi\goto -\varphi$ and the equality
$\langle 0| e^{i\alpha\varphi}|0\rangle =
\langle 0|e^{-i\alpha\varphi}|0\rangle$ follows.} 
\EQ
-i\,{\mbox Res}_{\th=\pm i\pi}F_{s\bar{s}}^\Psi(\th) = 
\left[\cos\delta-\cos(\delta\mp 2\pi\alpha/\beta)\right]\,
\langle 0|e^{i\alpha\varphi}|0\rangle\,\,\,.
\label{respsi}
\EN

The presence of these poles has a drastic consequence on the spectrum 
of the solitonic sector of the perturbed theory. Indeed, for generic 
values of $\alpha$ and $\delta$, the non--zero residue (\ref{respsi}) 
implies an infinite correction to the soliton and antisoliton masses. 
This divergence may be viewed as the technical manifestation of the fact 
that soliton and antisoliton no longer survive as asymptotic particles 
of the perturbed  theory. Of course, this result could be anticipated by
observing that for generic values of $\alpha$, the lagrangian
(\ref{lagrangian}) loses its original $2\pi/\beta$-periodicity, so that 
solitons and antisolitons become unstable excitations. As discussed
in more detail in Appendix\,B, the lifting of the degeneracy of the 
original minima induces a linear attractive potential between the
soliton-antisoliton pairs which causes their collapse into a string 
of bound states. 

Let us consider more closely the situation in which $\omega = m/n$, 
with $m$ and $n$ coprime integers, so that the lagrangian
(\ref{lagrangian}) retains a $2\pi n/\beta$-periodicity. As just
argued, as soon as the perturbation $\lambda \cos(\alpha 
\varphi + \delta)$ is switched on, the solitons of the original theory
become confined into states with zero topological charge\footnote{This 
is a statement which holds or $n \neq 1$  and generic values of 
$\delta$.} but in this case ``packets'' formed by $n$ of the original 
solitons (or antisolitons) survive as stable excitations.
These are nothing else but the topological excitations which interpolate
between the degenerate minima of the perturbed potential\footnote{Such 
packets, considered as a whole, are local with respect to the exponential 
operator $e^{i\alpha\varphi}$ since their semi-locality phase is 
$e^{\pm 2i\pi n\alpha/\beta} = 1$.} (see Figure 1.c and 2.b). Clearly, a 
complete space of topological excitations for the perturbed theory 
can be constructed  in terms of the ``$n$-soliton'' and 
``$n$-antisoliton'' states. 

The above conclusion holds for generic values of $\delta$. However, 
for specific values of this parameter, the single-soliton states of
the unperturbed theory may remain stable and unconfined even in the 
presence of the perturbation. In the FFPT framework, this corresponds 
to the case in which the correction to the soliton mass is finite, namely 
when the residue on the annihilation pole (\ref{respsi}) vanishes.
Taking into account the condition (\ref{3request}) satisfied by $\delta$ 
in the rational case, the residue (\ref{respsi}) vanishes when 
\EQ
|\delta| =\pi/n \,\,\,
\label{resvan1}
\EN
provided $m$ and $n$ satisfy the condition 
\EQ 
|k n - m| = 1 \,\,\, {\makebox{\rm for some}} \,\,k\in{\bf Z}\,\,\,.
\label{resvan2}
\EN
These are precisely the values of $\omega$ and $\delta$ for which 
the minimum at $\varphi = 0$ of the unperturbed potential becomes 
degenerate with one of its adjacent minima when the perturbation is 
switched on (Figure 3.b), and therefore the single-soliton
excitations interpolating between the two minima remain stable. Note, 
however that it is generally impossible to simultaneously cancel 
both the annihilation poles at $\th = i\pi$ and $\th = -i\pi$; 
it can be argued that this corresponds to the fact that only one of the 
adjacent minima (the one located at $\varphi = 2\pi/\beta +
O(\eta)$ or the one located at $\varphi = -2\pi/\beta + O(\eta)$) 
can become degenerate with the absolute minimun around 
$\varphi = 0$ (order $\eta$). Suppose that the
degeneracy is realised between the central minimum and its right 
neighbour. The presence of the degeneracy implies that, for the 
values of the parameters we are considering, the $n$-soliton discussed 
above actually breaks into the sequence of two asymptotically stable 
excitations: a kink $K_1$ of mass $M$ interpolating between $\varphi = 0$ 
and $\varphi = 2\pi/\beta$, and a kink $K_{n-1}$ of mass $(n-1) M$
interpolating between $\varphi = 2\pi/\beta$ and $\varphi = 2\pi n/\beta$
(here and below $O(\eta)$ corrections to the masses and positions of the 
minima are understood). Obviously, this pattern is repeated with periodicity
$2\pi n/\beta$.

More generally, it is not difficult to see that under the action of the
perturbation, for $\delta = \varepsilon\pi/n$ ($\varepsilon = \pm 1$), and 
for any pair $m$,$n$ ($m \neq 1$), the absolute minimum close to $\varphi =
0$ remains degenerate  with the minimun located near to $\varphi =  
2\pi j/\beta$, with $j$ given by 
\EQ
j\, =\, \frac{k n -\varepsilon}{m} \in \{1,2,\ldots,n-1\}\,\,\,.
\label{rational}
\EN
There is always a unique integer $k$ for which the 
above equation is satisfied. In this case, the $n$-soliton breaks 
into a kink $K_j$ with mass $j M$ and a kink $K_{n-j}$ with mass 
$(n-j) M$.

The denomination of kinks we gave to these excitations requires some 
clarification. Generally, kinks are elementary excitations which
interpolate between different vacua of a theory with spontaneously 
broken symmetry. When the structure of the vacua is sufficiently 
non-trivial, suitable ordering prescriptions must be assigned to
construct acceptable multi-kink configurations. In this circumstance, 
kinks drastically differ from ordinary particles, in terms of which the 
space of multi-particle states can be constructed without any restriction. 
The situation we are dealing with is not of this kind. In fact, due to 
the simple vacuum structure (alternation of two different kinds of minima), 
any string of $K_j$ and $K_{n-j}$ is an allowed multi-kink configuration. 
Differently stated, for any sequence of $K_j$ and $K_{n-j}$ there exists 
an unique interpolation pattern among the degenerate vacua of the
theory\footnote{Since we are dealing with excitations above the real 
vacuum of the theory, the interpolation starts from $\varphi=0$.
Moreover, the sequences $K_j K_j$ and $K_{n-j} K_{n-j}$ can only 
correspond to going back and forth between a couple of consecutive 
minima. Then, as long as the alternating minima remain inequivalent, 
there is no need to define antikink states.}.

It should be clear that the above considerations regarding the
evolution of the particle spectrum of the theory (\ref{actiondsg}) 
have a validity which goes beyond the perturbative framework in which 
they have been derived: the stable or instable character of an excitation 
under the action of the perturbation is a qualitative feature which can be 
extended away from the perturbative domain. For example, had we 
performed our perturbative analysis in the limit $\eta \goto \infty$ 
rather than in the limit $\eta\goto 0$, we would have established the
existence of $m$-solitons rather than $n$-solitons. These are different 
names for an unique excitation which exists and is stable for any value of
$\eta$. Simply, it is intuitively helpful to see this object as the bound 
state of $n$ solitons of the $\eta = 0$ theory, in one perturbative limit, 
or $m$ solitons of the $\eta = \infty$ theory, in the other.

\subsection{Phase Transition}

Let us apply the above considerations to the case $\delta =\pm \pi/n$ in
which the $n$-solitons ``decay'' into the sequence of kinks $K_j$ and
$K_{n-j}$. By increasing $\eta$, the masses of these particles evolve
preserving though their stability, and for $\eta$ very large they can be 
reinterpreted perturbatively as $K_l$ and $K_{m-l}$ kinks, with $l$
determined by eq.\,(\ref{rational}) but with $m$ and $n$ interchanged. 

An exception to the above smooth evolution pattern arises when $m =1$: 
in this case, the $n$-soliton -- which may be seen as ``sum'' of 
two kinks $K_1$ and $K_{n-1}$ -- evolves for $\eta \goto \infty$ into a
``$1$-soliton''. The latter, up to a mass correction, is simply 
the soliton of the $\eta = \infty$ theory and therefore is completely stable
(in this limit the perturbation is local with respect to all the excitations 
of the unperturbed theory for $m=1$). Thus, a composite topological
excitation has surprisingly transmuted into an elementary one along
the way! 

Such a transmutation necessarily requires the existence of an 
intermediate critical value $\eta_{c}$ at which a phase transition 
takes place\footnote{One could imagine several critical thresholds. 
The simplest possibility is assumed here.}. The phase transition 
presents the following pattern. For $\eta < \eta_c$, the $n$-soliton
consists of the sequence of the two kinks $K_1$ and $K_{n-1}$. When 
$\eta \rightarrow \eta_c$ from below, the intermediate minimum which 
is in common between the two kinks approaches the one initially located 
at $\varphi = 0$ (Figure 5.a). The two minima become coincident at $\eta =
\eta_c$ (Figure 5.b) and, at this value, $K_1$ disappears as a topological
excitation. In particular, its mass (which is proportional to the potential 
barrier between the minima) shrinks from M at $\eta = 0$ to $0$ at 
$\eta = \eta_c$. For $\eta > \eta_c$ the original composite topological 
excitation consists of a single massive kink, which interpolates
between the vacua of the $\eta = \infty$ theory (Figure 5.c). 
However, the dynamical degree of freedom carried by $K_1$ do not 
disappear across the transition line: it is in fact transferred to a
particle with zero topological charge which is precisely massless at 
$\eta = \eta_c$ and becomes a breather at $\eta = \infty$\footnote{Note  
that the existence of at least one breather in this limit is ensured, 
for $m=1$, by the condition $\alpha \beta <4 \pi$ we derived in the 
previous section.}. This tranformation of a topological excitation into a
non-topological one should not be surprising: we already remarked that, 
from the point of view of the construction of the space of asymptotic 
quantum states, $K_1$ is actually indistinguishable from an ordinary 
particle.

The presence of a massless particle in the spectrum of the $\eta = \eta_c$ 
theory amounts to say that the correlation length is infinite along the RG 
trajectory labelled by $\eta_c$. The correlation length cannot decrease in
the RG flow towards the infrared limit, therefore such a trajectory must
end into a new fixed point. Since the theory (\ref{actiondsg}) is unitary, 
Zamolodchikov's C-theorem \cite{C-th} ensures that such a fixed point is 
described by a CFT with central charge smaller than that of the ultraviolet 
fixed point, namely $C=1$. On the other hand, it is known from 
Ref.\,\cite{FQS} that the only unitary CFTs with $C<1$ are those belonging 
to the unitary minimal series, with central charges $C_p = 1-6/[p(p+1)]$, 
$p=3,4,\ldots$. For $\eta$ slightly below $\eta_c$, the mass of $K_1$ is 
much smaller than the mass of the other particles of the theory. Thus, the 
low energy limit is effectively described only by the kink 
$K_1$ interpolating among the vacuum and its adjacent degenerate minimum. 
Stated in a Landau-Ginzburg language, the action (\ref{actiondsg}) 
can be effectively truncated to a spontaneously broken $\varphi^4$ theory, 
which is well known to describe the universality class of the Ising model. 
In fact, the one we described above is exactly the Ising phase transition 
with the parameter $\eta$ playing the role of the temperature. Then we
conclude that the RG trajectory labelled by $\eta_c$ interpolates between 
the gaussian fixed point with $C=1$ and the Ising fixed point with 
$C = C_3 = 1/2$ (Figure 6). A model which provides an
explicit realization of this flow is discussed in the next Section.

\resection{Two examples}

This section is intended to illustrate the wide applicability of the 
ideas discussed in this paper through a couple of examples referring 
to quite different physical situations. The first example is taken from 
classical statistical mechanics and deals with a deformation of the
Ashkin-Teller model. The second one comes from quantum field theory and
concerns the massive Schwinger model. Our interest is in exhibiting 
those features of the two models which are relevant for the discussion 
developed in this paper. The reader is referred to the original literature 
for comprehensive studies on both subjects (see Refs.\,\cite{Kadanoff,KB} 
for Ashkin-Teller and \cite{CJS,Coleman2} for the massive Schwinger model).

\subsection{The Generalized Ashkin-Teller Model}

The Ashkin-Teller model describes two planar Ising models interacting
through a local four spin interaction. It is defined by the lattice 
Hamiltonian
\EQ
H_{AT}\,=\,\sum_{(i,j)}[J(\sigma_1^i\sigma_1^j+\sigma_2^i\sigma_2^j)+
K\sigma_1^i\sigma_1^j\sigma_2^i\sigma_2^j]\,\,,
\label{H}
\EN
where $\sigma_{1,2}^i=\pm 1$, the sum is over nearest neighbours and the same 
coupling $J$ has been chosen for the two Ising models (isotropic case). 

A scaling limit description of the model can be formulated in terms of 
the euclidean action of a two--layer Ising system (Figure 7)
\EQ
{\cal A}_{AT} \,=\,{\cal A}_1^{Ising}+{\cal A}_2^{Ising}+
\tau\int d^2x\,(\varepsilon_1(x)+\varepsilon_2(x))+
\rho\int d^2x\,\varepsilon_1(x)\varepsilon_2(x)\,\,,
\label{at}
\EN
where ${\cal A}_i^{Ising}$ and $\varepsilon_i(x)$ denote the fixed 
point action and the energy operator of the $i$-th Ising model, 
respectively. In this language, the scaling limit of the Ashkin-Teller
model appears as a CFT with central charge $C = 1/2+1/2 = 1$
perturbed by the two operators ${\cal E} \equiv \varepsilon_1 +
\varepsilon_2$ and  $\varepsilon \equiv \varepsilon_1\varepsilon_2$. 

Let us keep at first the two Ising models at their critical temperature
($\tau=0$). Since the operator $\varepsilon$ is marginal 
($\Delta_{\varepsilon} = 1/2+1/2$), its addition to the action of the two
critical Ising models does not spoil criticality, but leads to a line of 
$C=1$ fixed points parameterised by the values of the coupling $\rho$. 
It is known \cite{Kadanoff,KB} that, while the spin fields $\sigma_1$ and
$\sigma_2$ retain along the critical line the conformal dimensions which 
they have at the decoupling point $\rho=0$ (i.e. $\Delta_\sigma=1/16$), the
dimensionalities of the total energy operator ${\cal E} = \varepsilon_1 
+ \varepsilon_2$ and of the so called ``polarisation'' operator 
$P = \sigma_1\sigma_2$ become instead $\rho$-dependent. Their ratio, 
however, remains the same as at the decoupling point 
\EQ
\frac{\Delta_P(\rho)}{\Delta_{\cal E}(\rho)}=
\frac{\Delta_P(0)}{\Delta_{\cal E}(0)}=
\frac{1/16+1/16}{1/2} = \frac{1}{4}\,\,\,.
\label{rapporto}
\EN
Thus, the action (\ref{at}) with $\tau\neq 0$ can be seen as a $C=1$ CFT
perturbed by the operator ${\cal E}$ with $\rho$-dependent conformal 
dimension, namely as a Sine-Gordon model (\ref{sg}) with the frequency 
$\beta$ determined by the relation 
\EQ
\Delta_{\cal E}(\rho)=\beta^2/8\pi\,\,. 
\label{betarho}
\EN
Note that, since $\Delta_{\cal E}(0)=1/2$, the decoupling point corresponds
to $\beta=\sqrt{4\pi}$. For this value of the coupling, the SG model is
equivalent to a free Dirac fermion (see (\ref{g})), as expected from the 
well known equivalence between the thermal Ising model and a free Majorana 
fermion.

Consider now a generalisation of the Ashkin-Teller model defined by the 
action
\EQ
{\cal A}_h \,=\, {\cal A}_{AT} + h \int d^2x\,\sigma_1(x)\sigma_2(x)\,\, ,
\label{ah}
\EN
with an additional spin-spin interaction between the two planar Ising
models. 

For $\tau = 0$, it can be shown that the action (\ref{ah}) defines an 
integrable massive field theory whose spectrum and scattering
amplitudes have been determined in \cite{LLM}. 

For $\tau \neq 0$, there is however the possibility to make a fine--tuning 
of the two parameters $\tau$ and $h$ in such a way that a massless RG flow 
is obtained. The infrared fixed point can be easily identified by means of
the following physical consideration. Let us consider first the
strong--coupling limit $h\goto\infty$: in this regime, the last term in
(\ref{ah}) forces the spins $\sigma_1$ and $\sigma_2$ to assume the same 
value on each site. Hence, the system results effectively reduced to a 
single Ising model. The reduced system is generally massive but, as any 
other Ising model, it may become critical provided the temperature 
of the two original layers is appropriately tuned to its critical value. 
It can be easily argued that the strong--coupling limit scenario just
described should also occur for finite values of $h$: the coupling $h$ 
tends to order the system (hence introducing a finite correlation length) 
but this tendency can be contrasted by a value of the temperature 
sufficiently high. In conclusion, for each value of $\rho$, there
should be a critical line in the $\tau h$-plane along which the system 
flows from the $C=1$ fixed point to the $C=1/2$ Ising fixed point. For 
the symmetries of the problem, the approach to the Ising fixed point must
occur along one of the irrelevant directions of the conformal family of 
the identity operator: in fact, the massless flow is along the line which 
separates the high-- and the low--temperature phases, hence the conformal
families of the magnetic and energy operators of the Ising model are both 
ruled out since the former is odd under the $Z_2$ spin symmetry while the 
latter is odd under the high--low temperature duality of the model. So,
only the conformal family of the identity operator is left.  

Let us finally rephrase the above discussion in the bosonic language. 
Since ${\cal E} \sim\cos\beta\varphi$, eq.\,(\ref{rapporto}) suggests the
identification $P\sim\cos(\beta\varphi/2+\delta_P)$, with $\delta_P$ to be
fixed. The model (\ref{ah}) then corresponds to the two--frequency 
SG model (\ref{actiondsg}), with the identification $\mu = \tau$,
$\lambda = h$, $\alpha = \beta/2$, $\delta = \delta_P$, with $\beta$
determined in terms of $\rho$ through (\ref{betarho}). We found in the
previous section that, for $\alpha/\beta = 1/n$, a flow to the Ising 
fixed point takes place in the $\lambda\mu$-plane if $|\delta| = \pi/n$.
This implies $|\delta_P| = \pi/2$, so that $P \sim \sin (\beta\varphi/2)$.

\subsection{Massive Schwinger model}

The massive Schwinger model describes quantum electrodynamics in two
dimensions. It is defined by the Lagrangian
\EQ
{\cal L}_{MSM}=-\frac{1}{4}F^{\mu\nu}F_{\mu\nu}+
\bar{\psi}(i\gamma^\mu\partial_\mu-m-e\gamma^\mu A_\mu)\psi\,\,.
\label{msm}
\EN
If the ``quark'' mass $m$ is set to zero, the theory reduces to the ordinary 
Schwinger model which is known to be equivalent to a free scalar field of 
mass $e/\sqrt{\pi}$ \cite{Schwinger,Lowenstein}. On the other hand, if we
set $e=0$ in (\ref{msm}) we are left with a theory of free massive fermions
which, as already remarked in section\,2, is equivalent to a pure SG model 
with $\beta=\sqrt{4\pi}$. Putting this information all together, one obtains 
the following bosonised form of the theory (\ref{msm}) \cite{CJS}
\EQ
{\cal L}_{Bose}=\frac{1}{2}(\partial_\mu\varphi)^2-
\frac{e^2}{2\pi}(\varphi+\gamma)^2+cm^2\cos\sqrt{4\pi}\varphi\,\,,
\label{bose}
\EN
where $c$ is a positive constant whose precise value is immaterial for our 
considerations. The parameter $\gamma$ accounts for the fact
that, contrary to what happens in the two limiting cases $m=0$ and $e=0$, a 
shift of the field $\varphi$ does not leave invariant the complete theory 
(\ref{bose}): it is in fact proportional to the $\Theta$ angle of the gauge
lagrangian (\ref{msm}) which physically corresponds to the presence of a
constant background electric field which breaks parity \cite{Coleman2}. 

The bosonic lagrangian (\ref{bose}) can be seen as the specialisation of 
the theory (\ref{actiondsg}) to the case $\beta = \sqrt{4\pi}$, $\lambda 
= (e/\sqrt{\pi}\alpha)^2$, $\delta = \alpha\gamma$, in the limit $\alpha
\goto 0$. The analysis performed in the previous sections for the theory 
(\ref{actiondsg}) is easily adapted to the present case. The requirement 
that the lagrangian (\ref{bose}) describes theories which do not differ 
simply for a shift of the vacuum leads to the restriction $|\gamma|\leq
\sqrt{\pi}/2$. Chosing the $\Theta$ angle to take values in the interval 
$(-\pi,\pi)$, then one has $\gamma = \Theta/\sqrt{4\pi}$. The potential 
term of the lagrangian (\ref{bose}) for a generic value of $\Theta$ and 
for the special value $\Theta = \pi$ is drawn in Figure 8. 

Let us apply the perturbative arguments of Section\,4 to the limit 
$e/m\goto 0$ of (\ref{bose}). The unperturbed theory is the SG at the 
free fermion point. The spectrum consists of soliton and antisoliton only,
which in the fermionic description must be identified with the
non-interacting quark and antiquark. The perturbing operator $\Psi 
\equiv (\varphi + \gamma)^2$ is non-local with respect to $s$ and
$\bar{s}$. Such a non-locality is of a less simple form than that of the 
exponential operators $e^{i\alpha \varphi}$, but is however easily deduced
from the latter. In fact, taking the $k$-th derivative with respect to 
$\alpha$ of eqs.\,(\ref{crossing}) and (\ref{residue}) and then setting 
$\alpha=0$, one finds
\EQ
\langle 0|\varphi^k(0)|s(\theta_1+2i\pi),\bar{s}(\theta_2)\rangle=
\langle 0|(\varphi(0)-2\pi/\beta)^k|\bar{s}(\theta_2),s(\theta_1)\rangle\,\,,
\EN
\EQ
-i\,\mbox{Res}_{\theta_1-\theta_2=\pm i\pi}
\langle 0|\varphi^k(0)|s(\theta_1),\bar{s}(\theta_2)\rangle=
\langle 0|\varphi^k(0)-(\varphi(0)\mp 2\pi/\beta)^k|0\rangle\,\,.
\label{power}
\EN
With $\beta=\sqrt{4\pi}$ and $\langle 0|\varphi|0\rangle=0$, the last
equation gives
\EQ
i\,\mbox{Res}_{\theta=\pm i\pi}F_{s\bar{s}}^\Psi(\theta)=\pi\mp\Theta\,\,.
\label{nice}
\EN
The interpretation of this result follows from the general discussion of 
Section\,4. For $|\Theta|<\pi$ the soliton and antisoliton masses receive 
an infinite correction. In the fermionic description this amounts to say that
quarks and antiquarks are confined by an arbitrary small electromagnetic 
interaction. Quark-antiquark pairs are bounded by a linear potential and give
rise to a string of electrically neutral bound states. The confining 
potential and an estimate of the number of bound states as a function of 
energy are given by the formulae (\ref{potential}) and (\ref{nonrel}) with
$A=e^2/2$ and $B=\Theta/\pi$.

For $|\Theta|=\pi$, however, the annihilation pole in the soliton-antisoliton 
form factor of the perturbing operator is (partially) cancelled and the mass
correction is finite. Quark and antiquark survive unconfined in the perturbed 
theory as kinks interpolating between the vacuum located at $\varphi=
{\cal O}(e^2)$ and a degenerate adjacent minimum (the latter lies at 
$\varphi=\pm\sqrt{\pi}+{\cal O}(e^2)$ depending on the sign of $\Theta$) 
(Figure 8.b). Since no topological excitation is present in the opposite 
limit $e/m\goto \infty$, there must exist a critical value of the ratio 
$e/m$ for which a confining phase transition takes place. The arguments of
the previous section apply basically unchanged to the present situation and
one concludes that along the critical trajectory in the $em$-plane the system 
flows from the $C=1$ to the $C=1/2$ fixed point.

\resection{Conclusion}

 The multi--frequency Sine--Gordon model is an interesting example of 
non--integrable QFT. In this paper we have exploited the Form Factor 
Perturbation Theory to study the evolution of the particle spectrum 
of this model as a function of the parameter $\eta$ which labels its 
interpolation between two integrable SG models. The soliton sector is 
the one which results the most affected by the non--integrable dynamics: 
as discussed in Section 4 and also illustrated by the examples in Section 
5, its evolution strongly depends on the values assumed by the frequency 
ratio $\omega$ and the phase--shift $\delta$. In the FFPT, disappearing of
soliton states and confinement of their multi--particle excitations 
originates from the non--locality of the perturbation operator with respect
to  the field which creates the one--soliton state. In the special cases 
$\omega =1/n$ and $\delta = \pm 1/n$, a change of the vacuum state 
takes place by varying $\eta$, with the consequent presence of phase 
transition lines: these are generally associated to massless field 
theories which interpolate between the conformal field theory with $C=1$ 
and the one associated to the Ising model, with $C = 1/2$.   

It would be highly interesting to match the rich scenario of the 
multi--frequency SG model as coming from the analysis of its particle 
spectrum with a more detailed investigation of its dynamics, in 
particular the calculation of its scattering amplitudes. As shown 
in Ref.\,\cite{CPS} for the double SG model, a remarkable pattern of
resonances entering the collision events of the kinks already emerges 
at the classical level. A challenging open problem is therefore to see 
how the results of Ref.\,\cite{CPS} can be translated at the quantum 
level and conversion amplitudes between different types of kinks, 
inelastic events and energy levels of the resonance states of the 
non--integrable model can be computed.

\vspace{15mm} {\em Acknowledgements}. We wish to thank J.L. Cardy, M.
Fabrizio, A. Neveu, P. Simonetti and Al.B. Zamolodchikov for helpful 
discussions and M. Halpern for a useful remark. The authors would also 
like to thank the the Institute of Theoretical Physics in Santa Barbara 
and the organizers of the program {\em Quantum Field Theory in Low
Dimensions} for the warm hospitality during their staying at the institute, 
where this work was started.

\newpage

\appendix

\appsection

At the special values of the coupling $\xi=\pi/n$, $n=1,2,\ldots$, 
soliton-antisoliton scattering in the Sine-Gordon model becomes
reflectionless, and is described by the transmission amplitude
\EQ
S(\th)=-\prod_{k=1}^{n-1}\frac{\sinh\frac{1}{2}[\th+i\pi(1-k/n)]}
                              {\sinh\frac{1}{2}[\th-i\pi(1-k/n)]}\,\,,
\EN
in which the simple poles located at $\th=i\pi(1-k/n)$, $k=1,2,\ldots,n-1$, 
correspond to the breathers $B_k$ appearing as bound states in $s\bar{s}$ 
scattering.

As discussed in Section\,4, the soliton-antisoliton form factors of the 
exponential operators
\EQ
F_{s\bar{s}}^\alpha(\th_1-\th_2)\equiv\langle 0|e^{i\alpha\varphi(0)}|
s(\th_1)\bar{s}(\th_2)\rangle
\label{ff}
\EN
satisfy the unitarity and crossing equations
\EQ
F_{s\bar{s}}^\alpha(\th)=S(\th)F_{\bar{s}s}^\alpha(-\th)\,\,,
\label{unitarity}
\EN
\EQ
F_{s\bar{s}}^\alpha(\th+2i\pi)=e^{-2i\pi\alpha/\beta}
F_{\bar{s}s}^\alpha(-\th)\,\,.
\label{crossing}
\EN
Since we have chosen the vacuum of the theory to be located in $\varphi=0$, 
the reflection $\varphi\goto -\varphi$ maps $s$ in $\bar{s}$. The invariance
of the theory under this reflection then requires
\EQ
F_{s\bar{s}}^\alpha(\th)=F_{\bar{s}s}^{-\alpha}(\th)\,\,.
\label{parity}
\EN
We parametrise the matrix element (\ref{ff}) as
\EQ
F_{s\bar{s}}^\alpha(\th)=e^{-\alpha\th/\beta}\frac{P_\alpha(\th)}{\sinh n\th}
F_0(\th)\,\,,
\label{param}
\EN
where
\EQ
F_0(\th)=-i\sinh\frac{\th}{2}\exp\left[\int_0^\infty\frac{dx}{x}
\frac{\sinh\frac{x}{2}(1-\xi/\pi)}{\sinh\frac{x\xi}{2\pi}\cosh\frac{x}{2}}
\frac{\sin^2\frac{x(i\pi-\th)}{2\pi}}{\sinh x}\right]
\EN
satisfies
\bea
& & F_0(\th)=(-1)^{n+1}S(\th)F_0(\th)\,\,, \\
& & F_0(\th+2i\pi)=F_0(\th)\,\,.
\label{fmin}
\eea
The denominator $\sinh n\th$ in (\ref{param}) contains the kinematical pole 
in $\th=i\pi$ associated to the residue equation
\EQ
-i\,{\mbox Res}_{\th=i\pi}F_{s\bar{s}}^\alpha(\th)=
(1-e^{-2i\pi\alpha/\beta})\langle e^{i\alpha\varphi}\rangle\,\,,
\label{residue}
\EN
as well as the bound state poles associated to the residue equations
\EQ
-i\,{\mbox Res}_{\th=i\pi(1-k/n)}F_{s\bar{s}}^\alpha(\th)=
\Gamma_k\,F_k^\alpha
\label{bound}
\EN
(here $\Gamma_k=[-i\,{\mbox Res}_{\th=i\pi(1-k/n)}S(\th)]^{1/2}$
is the three-particle coupling between soliton, antisoliton and $k$-th 
breather and $F_k^\alpha\equiv \langle 0|e^{i\alpha\varphi(0)}|B_k\rangle$).
The pole in $\th=0$ is cancelled by the zero contained in $F_0(\th)$.

According to eqs.\,(\ref{unitarity}), (\ref{crossing}) and (\ref{parity}), 
$P_\alpha(\th)$ entering (\ref{param}) must satisfy
\bea
P_{\alpha}(\th)=(-1)^n P_{-\alpha}(-\th)\,\,,\\
P_{\alpha}(\th+2i\pi)=(-1)^{n+1}P_\alpha(\th)\,\,,
\eea
and then can be written as
\EQ
P_\alpha(\th)= \left\{
\begin{array}{ll}
\sum_{j=0}^N\left[a_j(\alpha)e^{j\th}-a_j(-\alpha)e^{-j\th}\right]\,\,, 
& \mbox{$n$ odd} \\
\sum_{j=0}^M\left[b_j(\alpha)e^{(j+1/2)\th}+b_j(-\alpha)e^{-(j+1/2)\th}
\right]\,\,, 
& \mbox{$n$ even.} \\
\end{array} \right.
\EN 
The integers $N$ and $M$ can be fixed from the asymptotic behaviour of the 
known solutions for
$\alpha=\pm \beta$ and $\alpha=\pm\beta/2$ \cite{Smirnov}. 
One finds $N=(n-1)/2$ and $M=(n-2)/2$. It is convenient to define 
$\tilde{a}_0(\alpha)=a_0(\alpha)-a_0(-\alpha)$, 
$c_j(\alpha)=a_j(\alpha)/\tilde{a}_0(\alpha)$, 
$d_j(\alpha)=b_j(\alpha)/b_0(-\alpha)$ and to rewrite $P_\alpha(\th)$ in the 
form
\EQ
P_\alpha(\th)= \left\{
\begin{array}{ll}
\tilde{a}_0(\alpha)\left[1+\sum_{j=1}^{(n-1)/2}\left(c_j(\alpha)e^{j\th}+
c_j(-\alpha)e^{-j\th}\right)\right]\,\,, 
& \mbox{$n$ odd} \\
b_0(\alpha)\sum_{j=0}^{(n-2)/2}\left[\frac{d_j(\alpha)}{d_0(\alpha)}
e^{(j+1/2)\th}+d_j(-\alpha)e^{-(j+1/2)\th}
\right]\,\,, 
& \mbox{$n$ even} \\
\end{array} \right.
\label{last}
\EN 
in which an inessential overall normalisation has been made explicit and 
(both for $n$ odd and $n$ even) $n-1$ $\alpha$-dependent coefficients  
appear as the only remaining unknowns in the problem. They are fixed 
exploiting 
the following observation. The lightest breather $B_1$, being the particle 
interpolated by the elementary field $\varphi$, changes sign under the 
reflection $\varphi\goto -\varphi$. The $k$-th breather can be seen as a 
bound state of $k$ breathers $B_1$, and then behaves as $B_k\goto (-1)^kB_k$ 
when $\varphi\goto -\varphi$. As a consequence we have
\EQ
F_k^\alpha=(-1)^kF_k^{-\alpha}\,\,,\hspace{1cm}k=1,2,\ldots,n-1.
\EN
Hence, remembering eq.\,(\ref{bound}) and defining 
$Q_\alpha(\th)=e^{-\alpha\th/\beta}P_\alpha(\th)$, we obtain
\EQ
Q_\alpha\left(i\frac{\pi k}{n}\right) = (-1)^k \,
Q_\alpha\left(-i\frac{\pi k}{n}\right)\,\,, \hspace{1cm}k=1,2,\ldots,n-1.
\EN
This is a set of $n-1$ linear equations which uniquely determine the 
coefficients in (\ref{last}). The solution is
\EQ
c_j(\alpha)=\prod_{m=1}^j\tan\frac{\pi}{2n}\left(m-\frac{n+1}{2}\right)
\cot\frac{\pi}{n}\left[\frac{\alpha}{\beta}+
\frac{1}{2}\left(\frac{n+1}{2}-m\right)
\right]\,\,,\hspace{.6cm}j=1,\ldots,\frac{n-1}{2}
\EN
for $n$ odd, and
\EQ
d_j(\alpha)=\prod_{m=0}^j\tan\frac{\pi}{2n}\left(m-\frac{n}{2}\right)
\cot\frac{\pi}{n}\left[\frac{\alpha}{\beta}+
\frac{1}{2}\left(\frac{n}{2}-m\right)
\right]\,\,,\hspace{.6cm}j=0,1,\ldots,\frac{n-2}{2}
\EN
for $n$ even.

\appsection

We discussed in Section\,4 how the topological excitations of SG model can
undergo a confinement phenomenon under the action of a perturbation. In this
Appendix we give a more quantitative description of the confinement pattern 
within the semiclassical approximation.

Consider at first a soliton-antisoliton ($s\bar{s}$) state in the 
pure SG model (\ref{sg}). Classically, it is associated to a configuration 
of the field $\varphi$ taking the vacuum value $\varphi=0$ from minus spatial 
infinity to a point $x_1(t)$ (the position of the soliton at time $t$) where 
it switches to the value $\varphi=2\pi/\beta$ of the right adjacent minimun 
of the potential; it keeps this value until the point $x_2(t)$ (location of
the antisoliton) where it switches again to $\varphi=0$. Similarly, the 
antisoliton-soliton ($\bar{s}s$) configuration corresponds to an interpolation 
from $\varphi=0$ to the left adjacent minimum $\varphi=-2\pi/\beta$, and 
back (see Figure 9).

Let us add to the SG potential the perturbing term 
$V(\varphi)=-\lambda\cos(\alpha\varphi+\delta)$ which lifts 
the degeneracy among the vacuum at $\varphi=0$ and its two adjacent 
minima\footnote{For the reasons discussed in Section\,3 the perturbation must 
be such that the central mininum remains the lowest among the three we are 
considering here. This amounts to require $A$ positive and $|B|\leq 1$ in
(\ref{potential}) below. Since the present considerations involve only a 
localised region of the potential, they lead to less severe restrictions on 
$\delta$ than those obtained in Section\,3.}. The two configurations
$s\bar{s}$ and $\bar{s}s$ described above are no longer stable since a 
surplus of energy equal, respectively, to $U_{s\bar{s}} = (x_2-x_1)
[V(2\pi/\beta)-V(0)]$ and $U_{\bar{s}s} = (x_1-x_2) [V(-2\pi/\beta)-V(0)]$
is associated to the intermediate plateau. The tendency of the system is 
to shrink in order to minimise the energy of the configuration: hence, the 
new term gives rise to the following attractive potential between the 
soliton and the antisoliton
\EQ
U(x) = \theta(x) U_{\bar{s}s} + \theta(-x) U_{s\bar{s}} = A (|x|-B x )\,\,,
\label{potential}
\EN
where $x\equiv x_1-x_2$, $\theta(x)$ is the step function and 
\EQ
A=2\lambda\sin^2\frac{\pi\alpha}{\beta}\cos\delta\,\,,\hspace{.5cm}
B=\cot\frac{\pi\alpha}{\beta}\tan\delta\,\,.
\label{ab}
\EN
Such a linearly rising potential confines the soliton-antisoliton pair and 
gives rise to a discrete spectrum of bound states. A continuum spectrum 
corresponding to asymptotically free motion is recovered only in the limiting 
case $|B|=1$. 

The number of bound states $N(E)$ as a function of energy can be obtained by 
the semi--classical formula
\EQ
N(E) = \frac{1}{2\pi}\oint p\,dx\,\,,
\label{bs}
\EN
where the integral is computed along the classical orbit. In a 
non-relativistic approximation, the above integral computed with
$p = [M(E-2M-U)]^{1/2}$ gives 
\EQ
N(E) = \frac{4\sqrt{M}(E-2M)^{3/2}}{3\pi A(1-B^2)}\,\,,
\label{nonrel}
\EN
where $M$ is the soliton mass and the rest energy $2M$ of the 
soliton-antisoliton pair has been included. The number of stable bound 
states is $N(E_T)$, $E_T$ being the value of the lowest threshold in the 
energy plane (Figure 10). $E_T$ equals $4M$ for $\beta^2\geq 4\pi$ and 
twice the mass $m_1$ of the lightest breather for $\beta^2<4\pi$. Comparing
with the exact solution expressed in terms of Airy function of the 
Schr\"{o}dinger equation with the potential (\ref{potential}), 
the semi--classical formula (\ref{nonrel}) provides excellent estimate 
of the energy levels (an error of $0.7\%$ for the first energy level, 
which is further reduced at $0.1\%$ for the second one). Therefore, for 
all practical purposes, we can consider formula (\ref{nonrel}) as an 
exact one. 

The computation of the integral (\ref{bs}) with $p = [(E-U)^2-M^2/4]^{1/2}$ 
required in the relativistic case cannot be performed exactly\footnote{It
is easy to see though that in this case, $N(E)\sim E^2$
for large energy.}. However, it can be checked numerically that the 
non-relativistic computation provides a reasonable estimate for $N(E_T)$.

The stable bound states originating from the threshold of
soliton--antisoliton configurations may influence the large asymptotic
behaviour of correlation functions. This is certanly the case for those
fields which were coupled in the unperturbed
system to the $s \overline s$ state. Let ${\cal O}$(x) be one of these 
fields. Neglecting states with higher number of particles and masses, 
the two-point function of the operator ${\cal O}(x)$ in the unperturbed 
system was given by 
\EQ
G(x) \,=\, \langle {\cal O}(x)\,{\cal O}(0) \rangle \sim
\int_{(2 M)^2}^{+\infty} \frac{d\mu^2}{2\pi} \,
\rho(\mu^2) K_0(\mu \mid x\mid)  
\EN 
with $K_0(r)$ the Bessel function and 
\EQ
\rho(p^2) = \int \frac{dp_s}{(2\pi) 2 E_s}\,\frac{dp_{\overline s}}{(2\pi) 
2 E_{\overline s}} \delta^2(p-p_s-p_{\overline s}) 
\mid \langle 0 \mid {\cal O}(0)\mid s \overline s \rangle \mid^2 
\EN
Hence, the leading behaviour for large $\mid x\mid$ in the unperturbed
case is given by 
\EQ
G(x) \sim D \,\, \sqrt{\frac{M}{\pi}} \frac{e^{-2 M \mid x\mid}}
{\mid x\mid^{\frac{3}{2} + 2 \sigma}}
\EN 
where $D$ and $\sigma$ are defined by the behaviour of the spectral
function $\rho(\mu^2)$ near the threshold $2 M$, 
$\rho(\mu^2) \sim D\,(\mu - 2 M)^{2\sigma}$. When the branch cut 
associated to the threshold of the state $s \overline s$ breaks down into 
the infinite sequence of bound states -- $N_b$ of which stable and whose 
energies are obtained by inverting eq.\,(\ref{nonrel}) -- the correlation
function becomes  
\EQ
G(x) \, \sim \, \frac{4}{3}\, 
\sum_{n=1}^{N_b} \frac{1}{n^{1/3}}\, 
E_n\,\rho[(E_n)^2] K_0[E_n \mid x\mid] \,\,\,. 
\EN 
Hence, the leading asymptotic behaviour for $\mid x\mid
\rightarrow \infty$ is determined by the lowest of them and we have 
\EQ
G(x) \, \sim \, \frac{4}{3} \sqrt{\frac{\pi}{4 M}}
E_1\, \rho[E_1^2] \frac{e^{-E_1 \mid x\mid}}{x^{1/2}} \,\,\,.
\EN

\newpage

\hs

{\bf Figure Caption}

\vspace{5mm}

\begin{description}
\item [Figure 1]. Topological excitations: (a) soliton state, (b)
anti--soliton state, (c) multisoliton state. 
\item [Figure 2]. (a) Perturbation of the SG potential (full curve) 
by another periodic interaction (dotted curve); (b) when $\mid \delta\mid > 
\frac{\pi}{n}$, the vacuum moves away from the origin. 
\item [Figure 3]. Examples of degeneracy of the vacuum at the origin 
with one of the other $(n-1)$ minima for $\delta = \frac{\pi}{n}$: (a)
$\omega =\frac{m}{n}$ ($m \neq 1$); (b) $\omega
=\frac{1}{n}$. 
\item [Figure 4]. Straight line (\ref{identification}) on the torus 
${\cal T}$ in the rational (a) and irrational case (b). The black point 
identifies the location of the minimum of the potential on the torus.  
\item [Figure 5]. Form of the potential near the phase transition point: 
(a) $\eta < \eta_c$; (b) $\eta = \eta_c$; (c) $\eta > \eta_c$. 
\item [Figure 6]. RG flows from $C = 1$ to $C = \frac{1}{2}$. 
\item [Figure 7]. The two--layer Ising model. 
\item [Figure 8]. Potential term in the massive 
Schwinger model: (a) $\Theta \neq \pi$; (b) $\Theta = \pi$. 
\item [Figure 9]. (a) Two--particle kink--antikink and (b)
antikink--kink states. 
\item [Figure 10]. Bound states of the linear potential confining the
soliton-antisoliton pairs. The levels above $E_T$ correspond to unstable 
particles. 
\end{description}

\pagestyle{empty}
\newpage
\begin{figure}
\null\vskip 3cm 
\centerline{
\psfig{figure=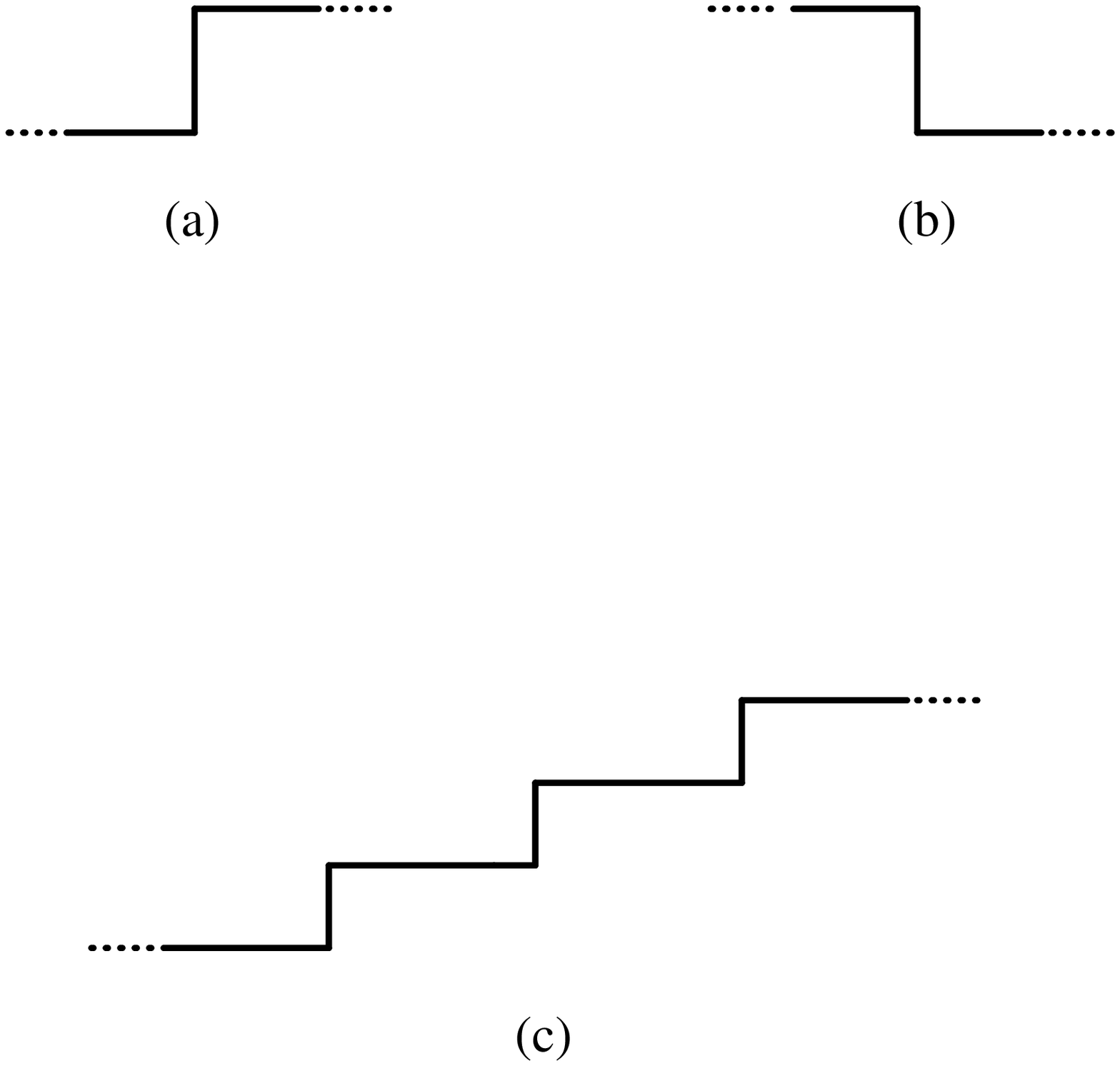}}
\vspace{1cm}
\begin{center}
{\bf \large{Figure 1}}
\end{center}
\end{figure}

\newpage
\begin{figure}[htb]
\hskip -45pt
\begin{minipage}[b]{.55\linewidth}
\centering\psfig{figure=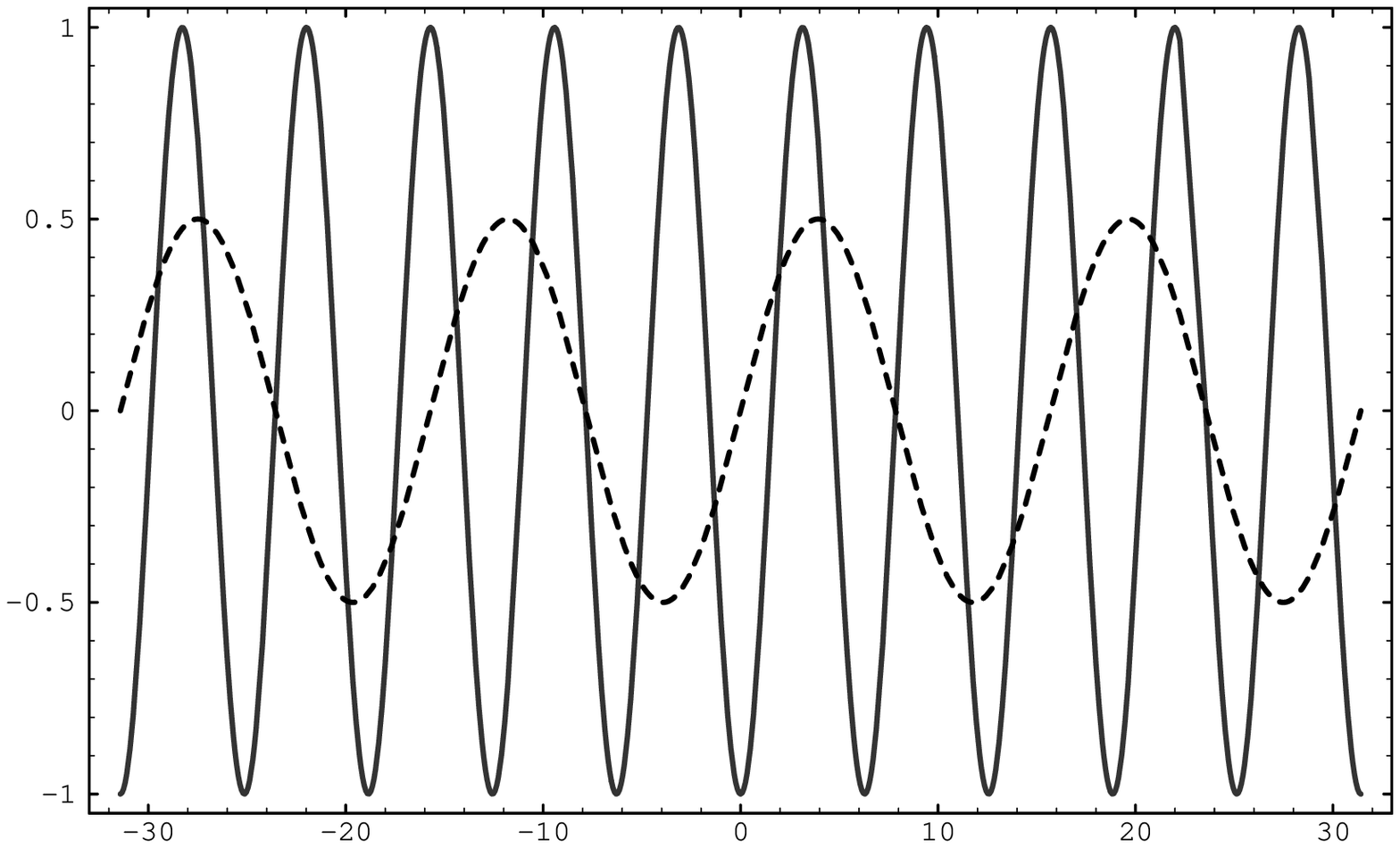,width=\linewidth}
\vspace{-25mm}
\begin{center}
{\bf (a)}
\end{center}
\end{minipage} \hskip 30pt
\begin{minipage}[b]{.55\linewidth}
\centering\psfig{figure=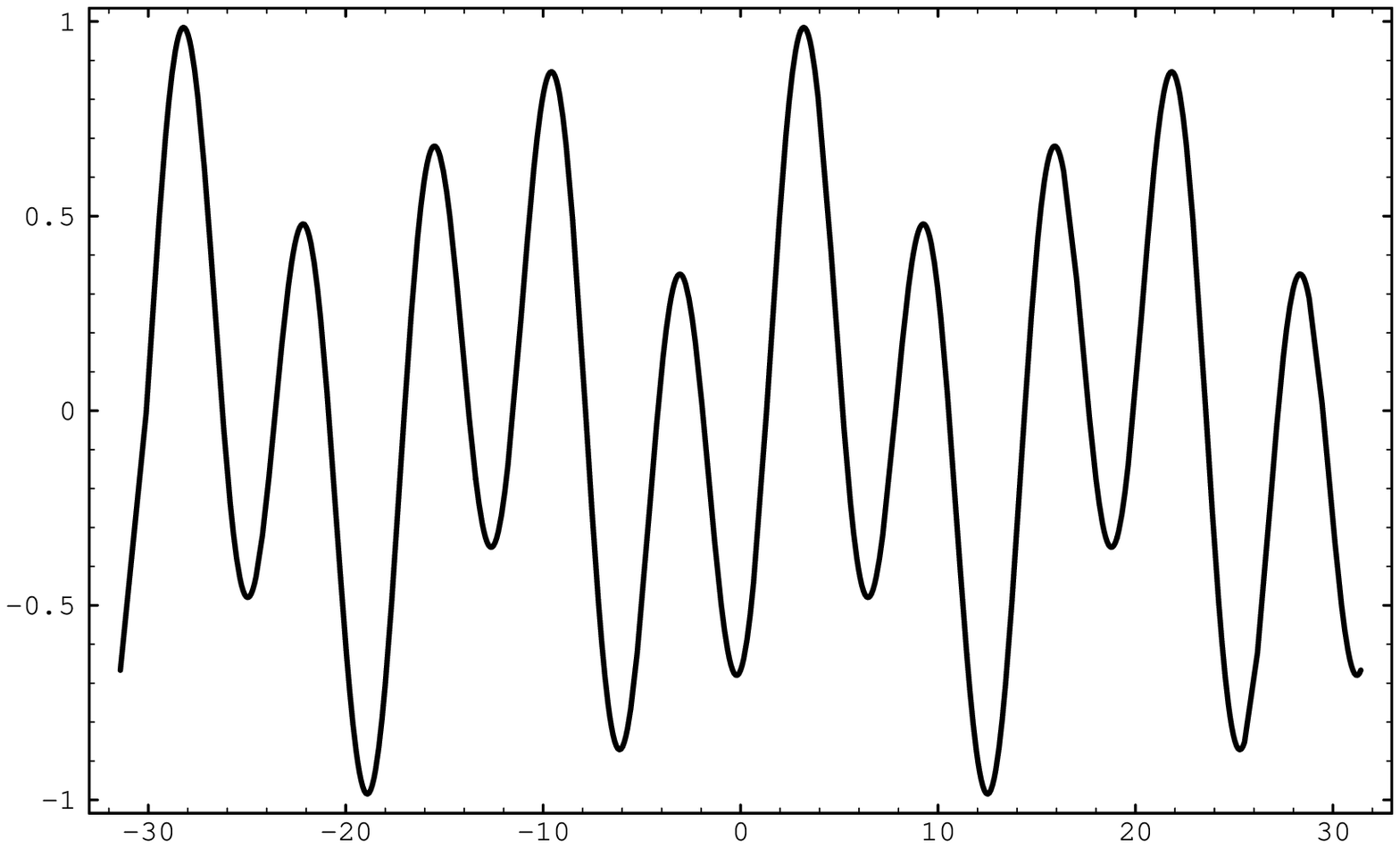,width=\linewidth}
\vspace{-25mm}
\begin{center}
{\bf (b)}
\end{center}
\end{minipage}
\vspace{15mm}
\begin{center}
{\bf Figure 2} 
\end{center}
\end{figure}

\newpage
\begin{figure}[htb]
\hskip -45pt
\begin{minipage}[b]{.55\linewidth}
\centering\psfig{figure=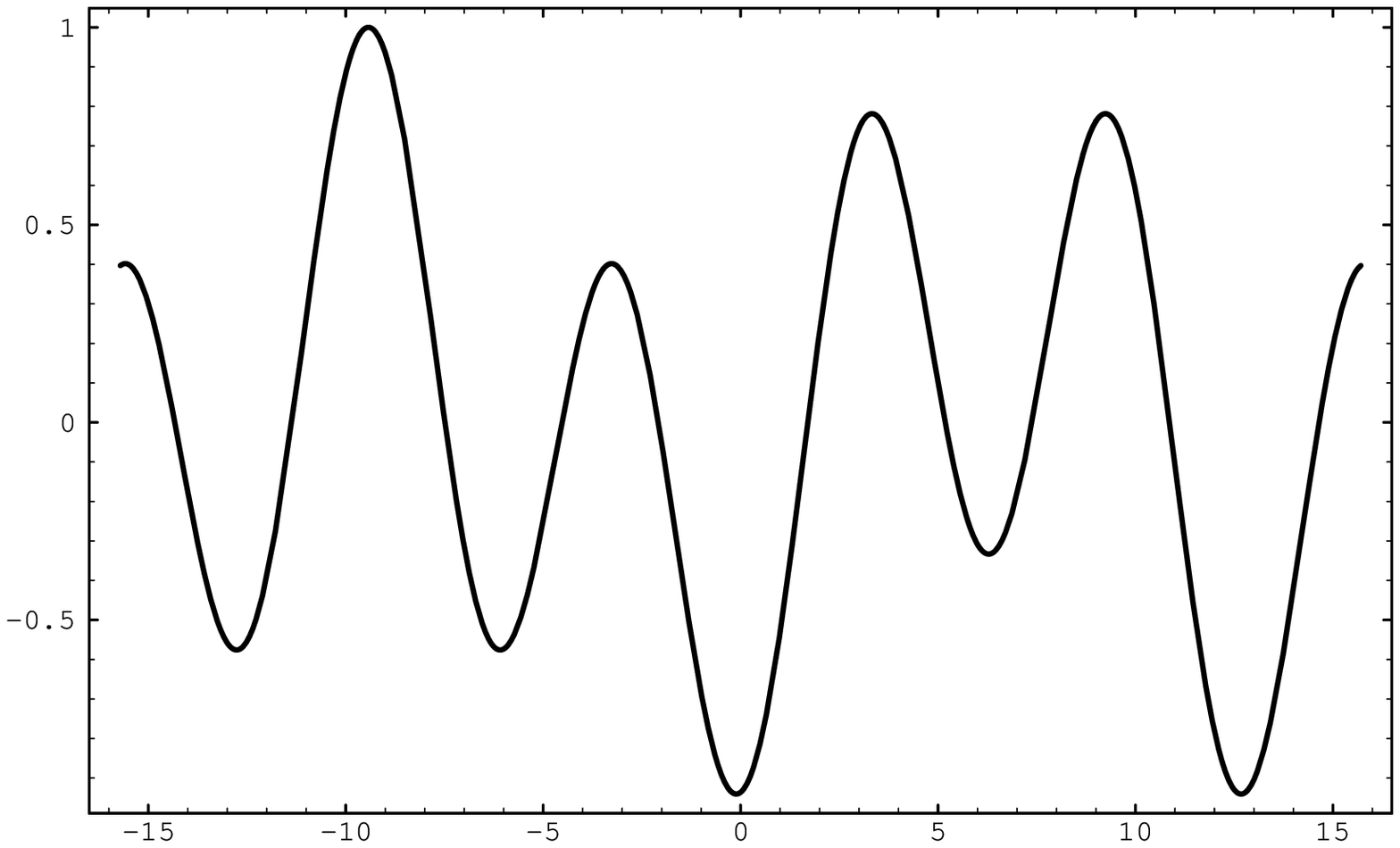,width=\linewidth}
\vspace{-25mm}
\begin{center}
{\bf (a)}
\end{center}
\end{minipage} \hskip 30pt
\begin{minipage}[b]{.55\linewidth}
\centering\psfig{figure=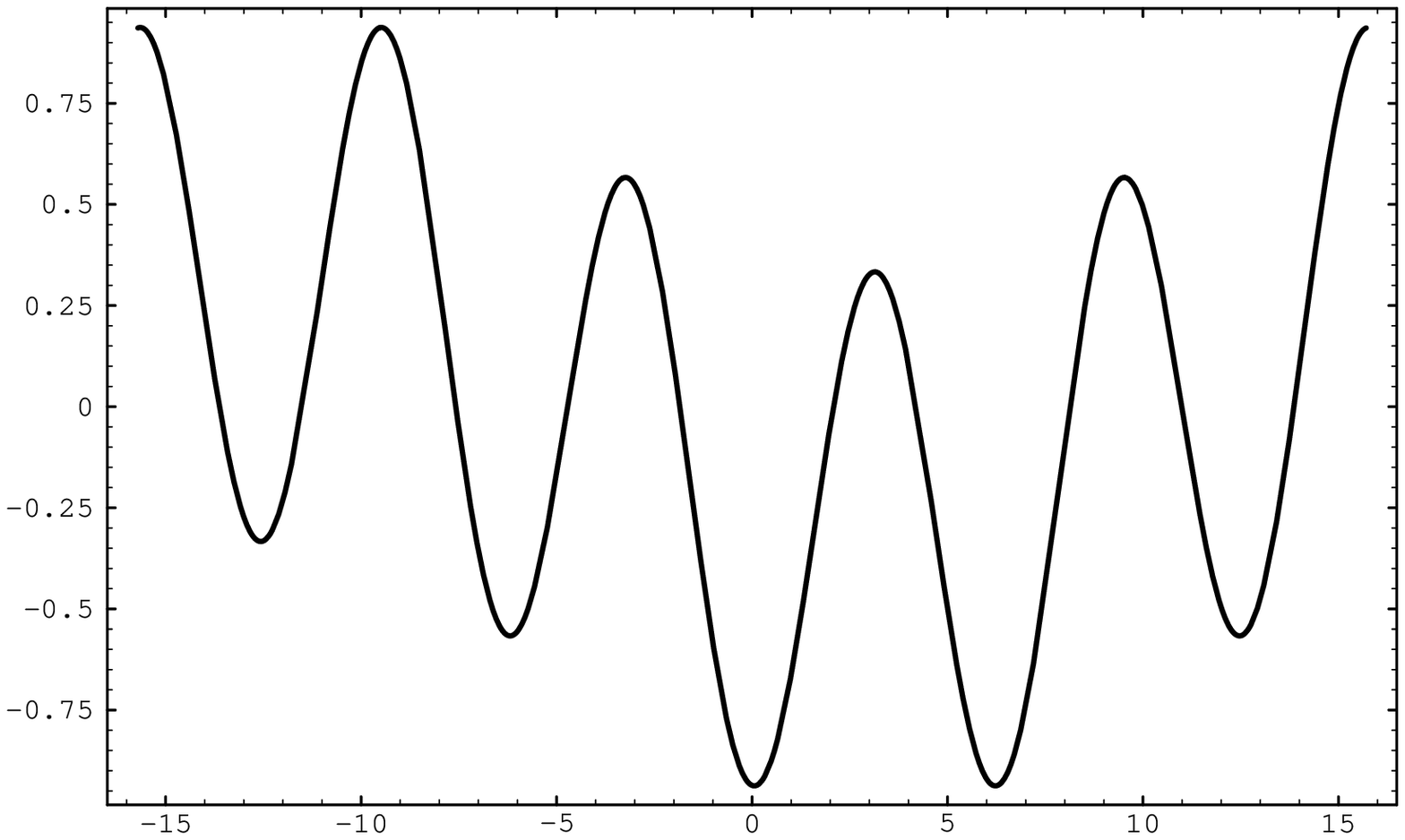,width=\linewidth}
\vspace{-25mm}
\begin{center}
{\bf (b)}
\end{center}
\end{minipage}
\vspace{15mm}
\begin{center}
{\bf Figure 3} 
\end{center}
\end{figure}

\newpage
\begin{figure}
\centerline{
\psfig{figure=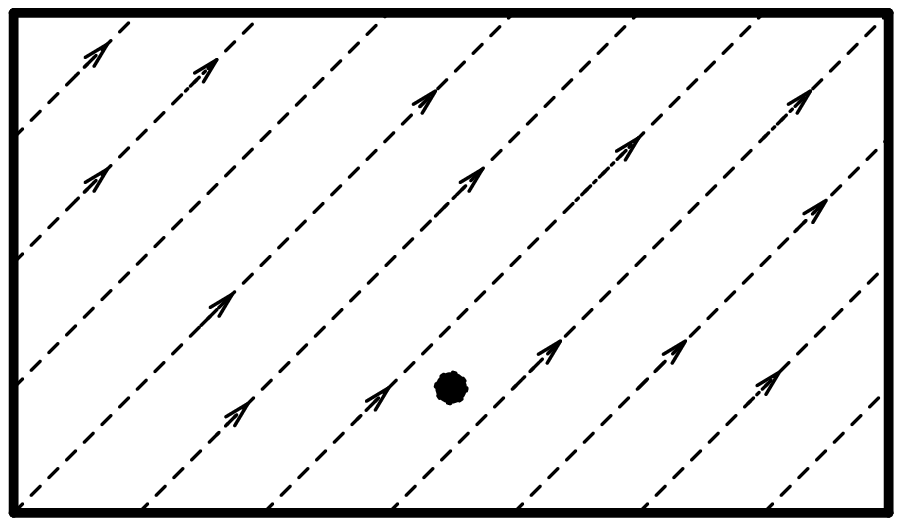}}
\vspace{0.5cm}
\begin{center}
{\bf (a)}
\end{center}
\vskip 1cm
\centerline{
\psfig{figure=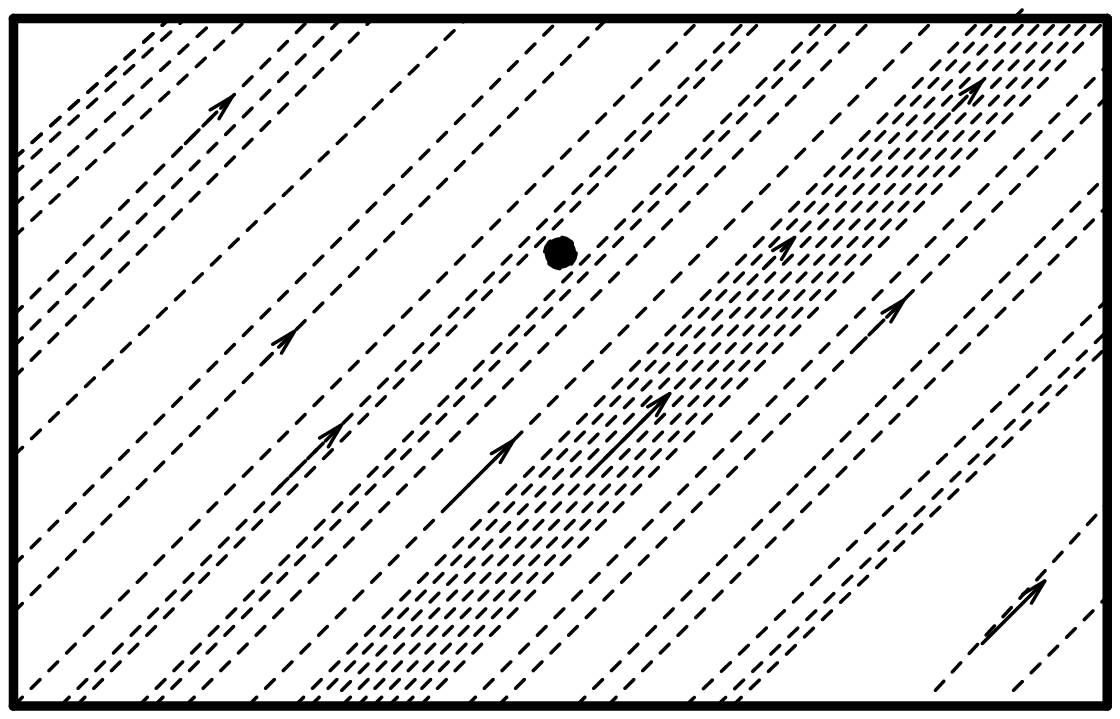}}
\vspace{0.5cm}
\begin{center}
{\bf (b)}
\end{center}
\vskip 1cm
\begin{center}
{\bf Figure 4}
\end{center}
\end{figure}

\newpage
\begin{figure}[htb]
\hskip -45pt
\begin{minipage}[b]{.55\linewidth}
\centering\psfig{figure=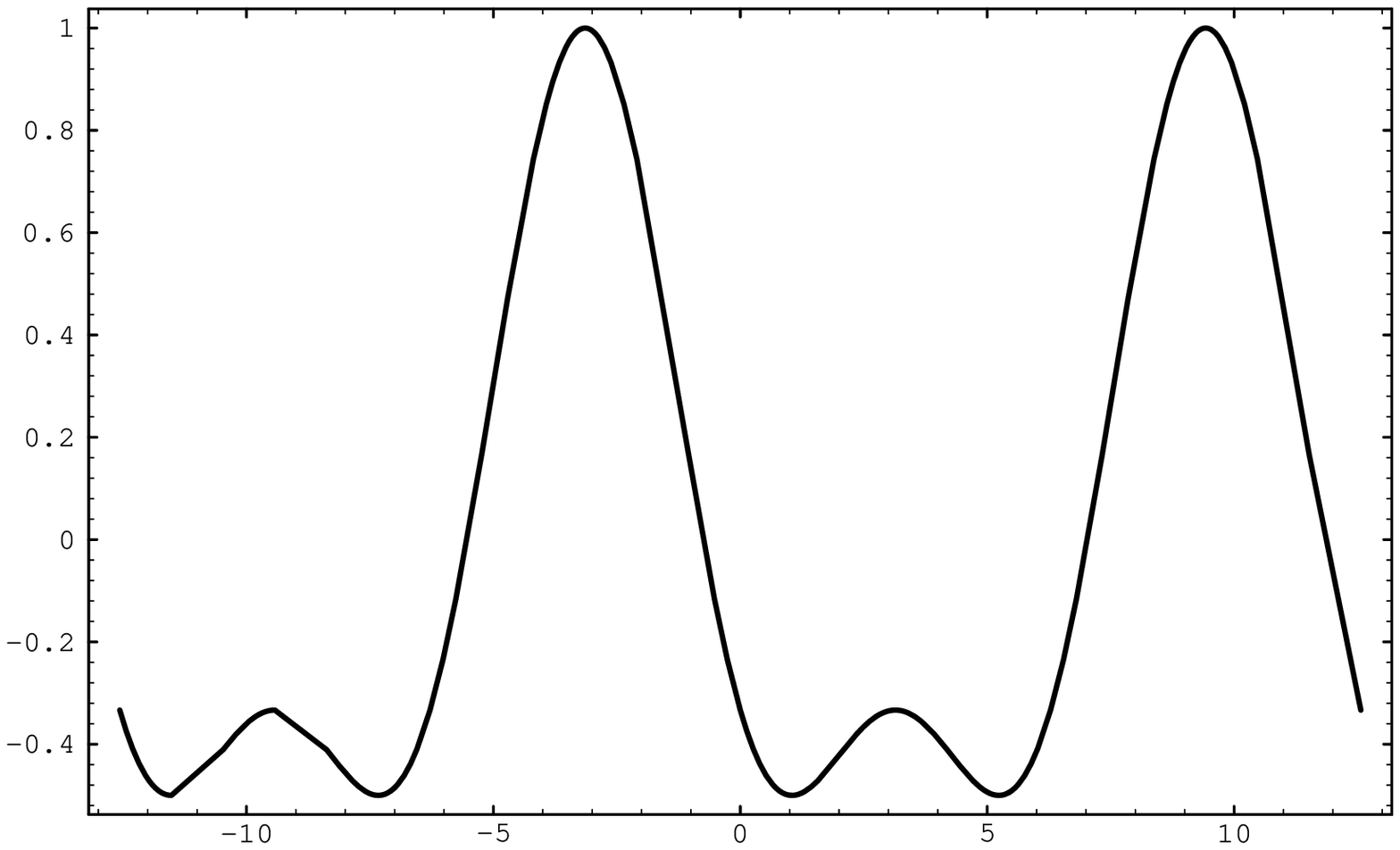,width=\linewidth}
\vspace{-25mm}
\begin{center}
{\bf (a)}
\end{center}
\end{minipage} \hskip 30pt
\begin{minipage}[b]{.55\linewidth}
\centering\psfig{figure=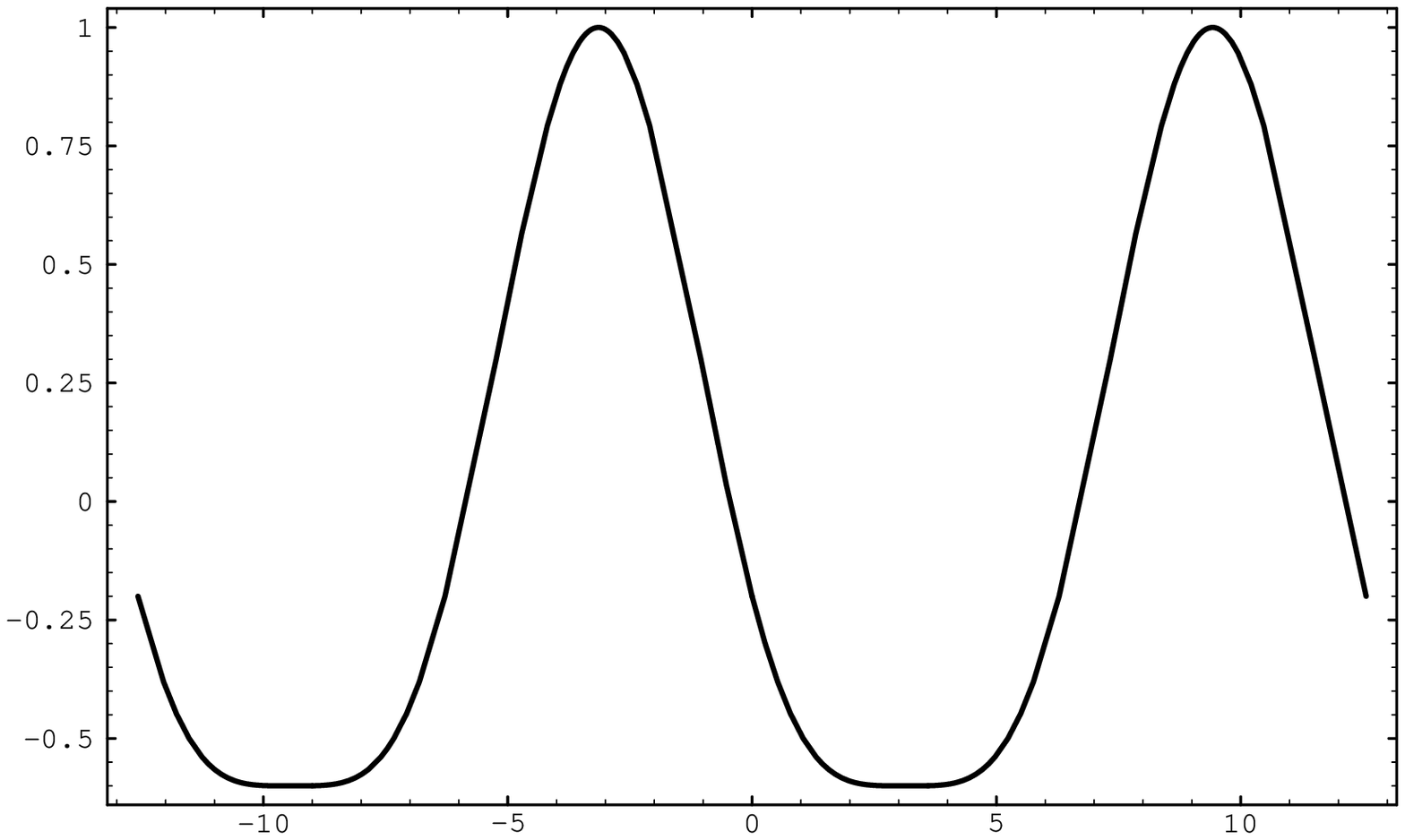,width=\linewidth}
\vspace{-25mm}
\begin{center}
{\bf (b)}
\end{center}
\end{minipage}
\begin{center}
\begin{minipage}[b]{.55\linewidth}
\centering\psfig{figure=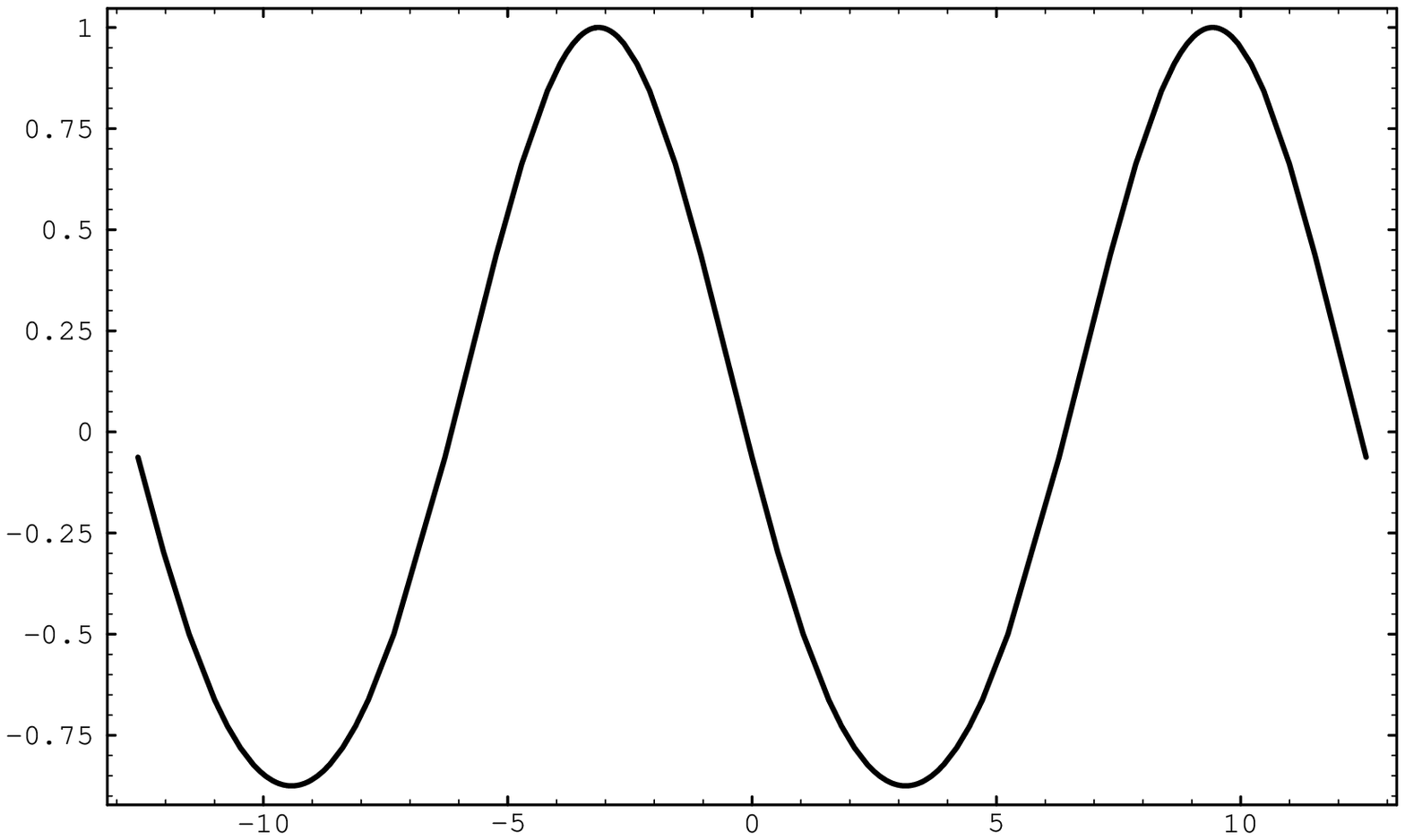,width=\linewidth}
\vspace{-25mm}
\begin{center}
{\bf (c)}
\end{center}
\end{minipage}
\vspace{25mm}
\begin{center}
{\bf Figure 5} 
\end{center}
\end{center}
\end{figure}

\newpage
\begin{figure}
\centerline{
\psfig{figure=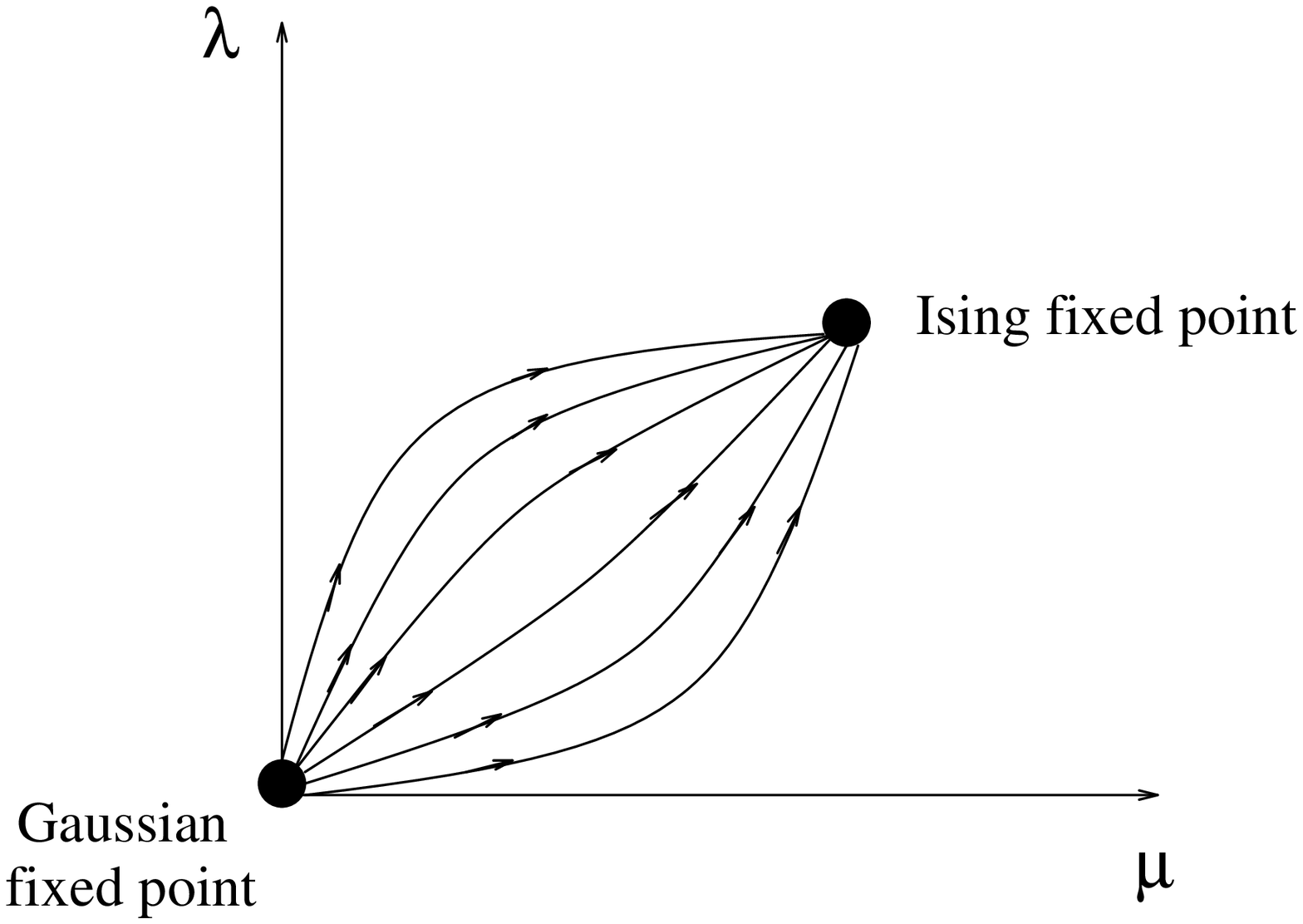}}
\vspace{5cm}
\begin{center}
{\bf Figure 6}
\end{center}
\end{figure}

\newpage
\begin{figure}
\centerline{
\psfig{figure=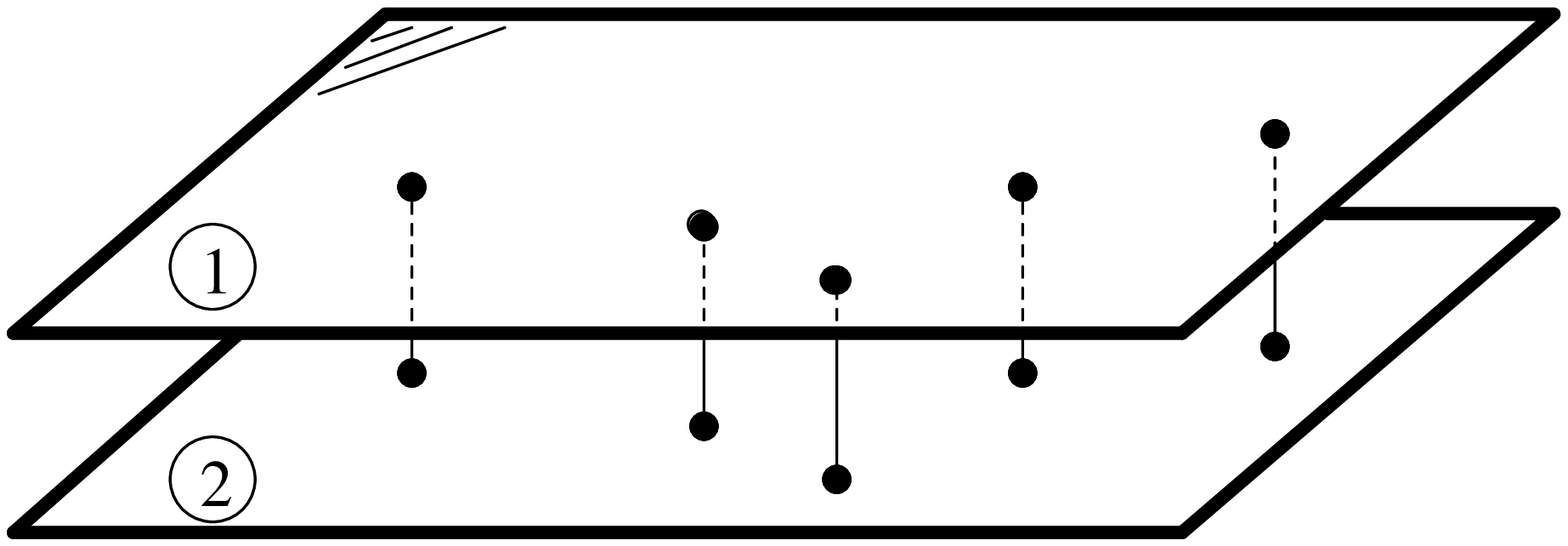}}
\vspace{5cm}
\begin{center}
{\bf Figure 7}
\end{center}
\end{figure}

\newpage
\begin{figure}[htb]
\hskip -45pt
\begin{minipage}[b]{.55\linewidth}
\centering\psfig{figure=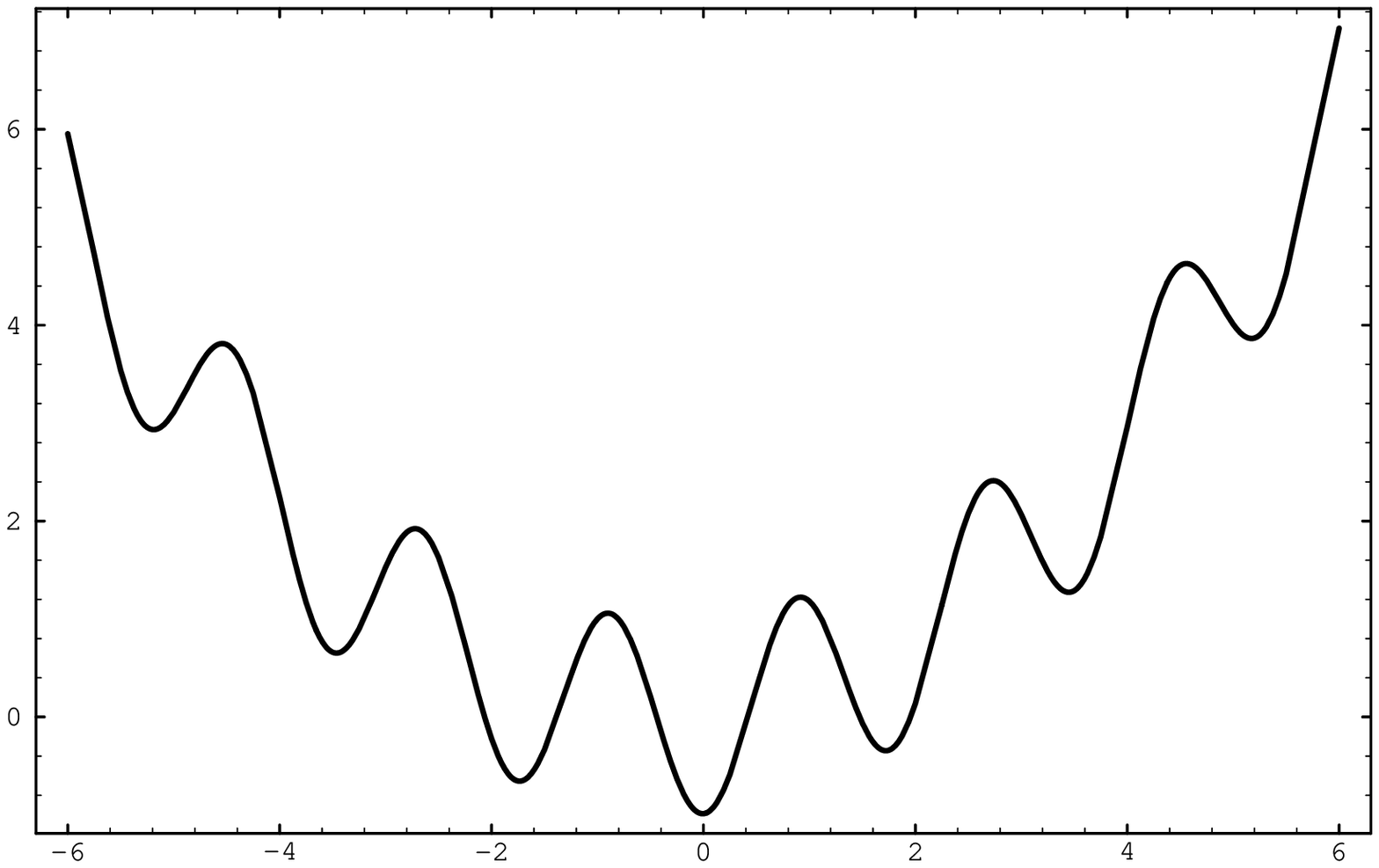,width=\linewidth}
\vspace{-25mm}
\begin{center}
{\bf (a)}
\end{center}
\end{minipage} \hskip 30pt
\begin{minipage}[b]{.55\linewidth}
\centering\psfig{figure=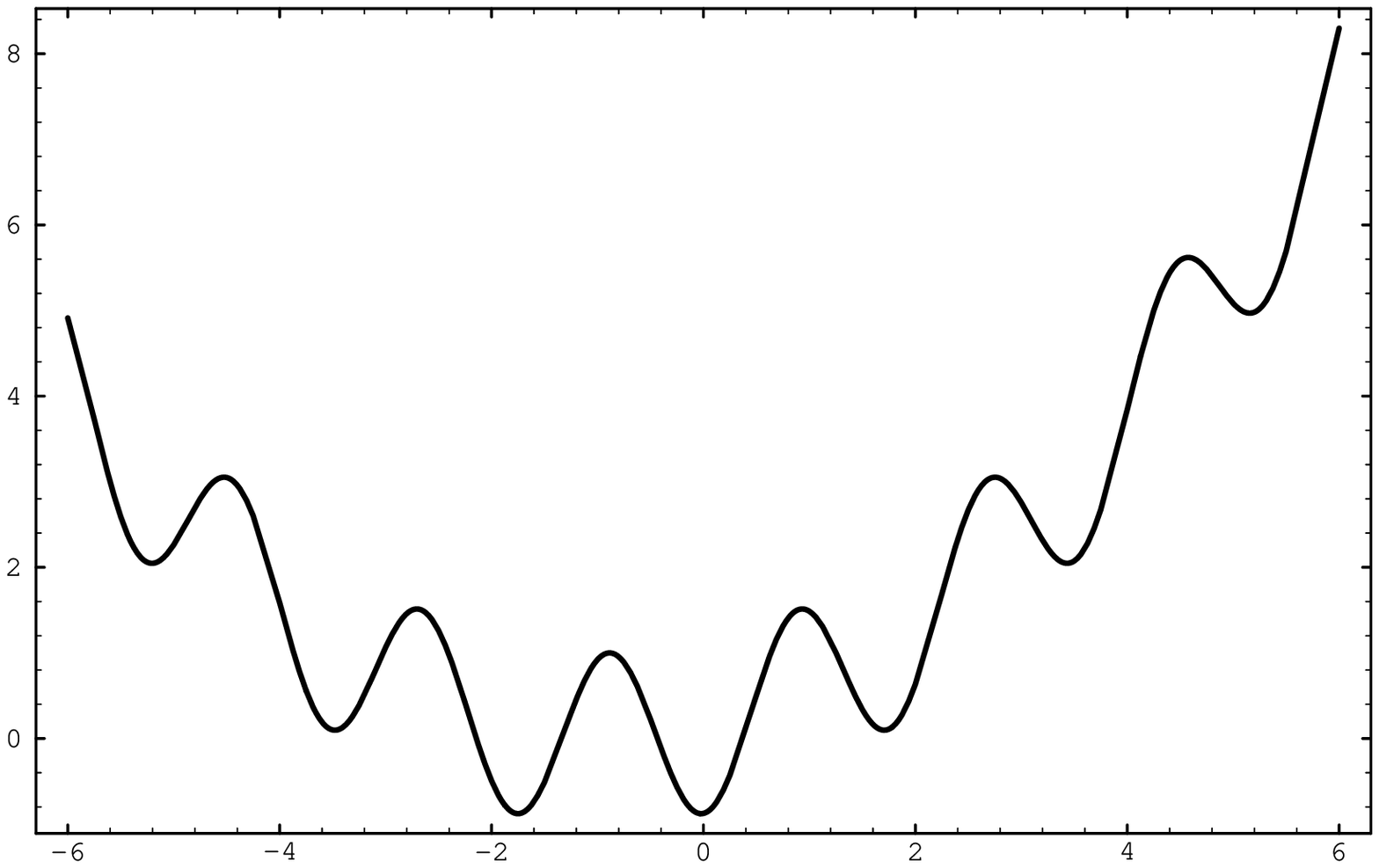,width=\linewidth}
\vspace{-25mm}
\begin{center}
{\bf (b)}
\end{center}
\end{minipage}
\vspace{25mm}
\begin{center}
{\bf Figure 8} 
\end{center}
\end{figure}

\newpage
\begin{figure}
\centerline{
\psfig{figure=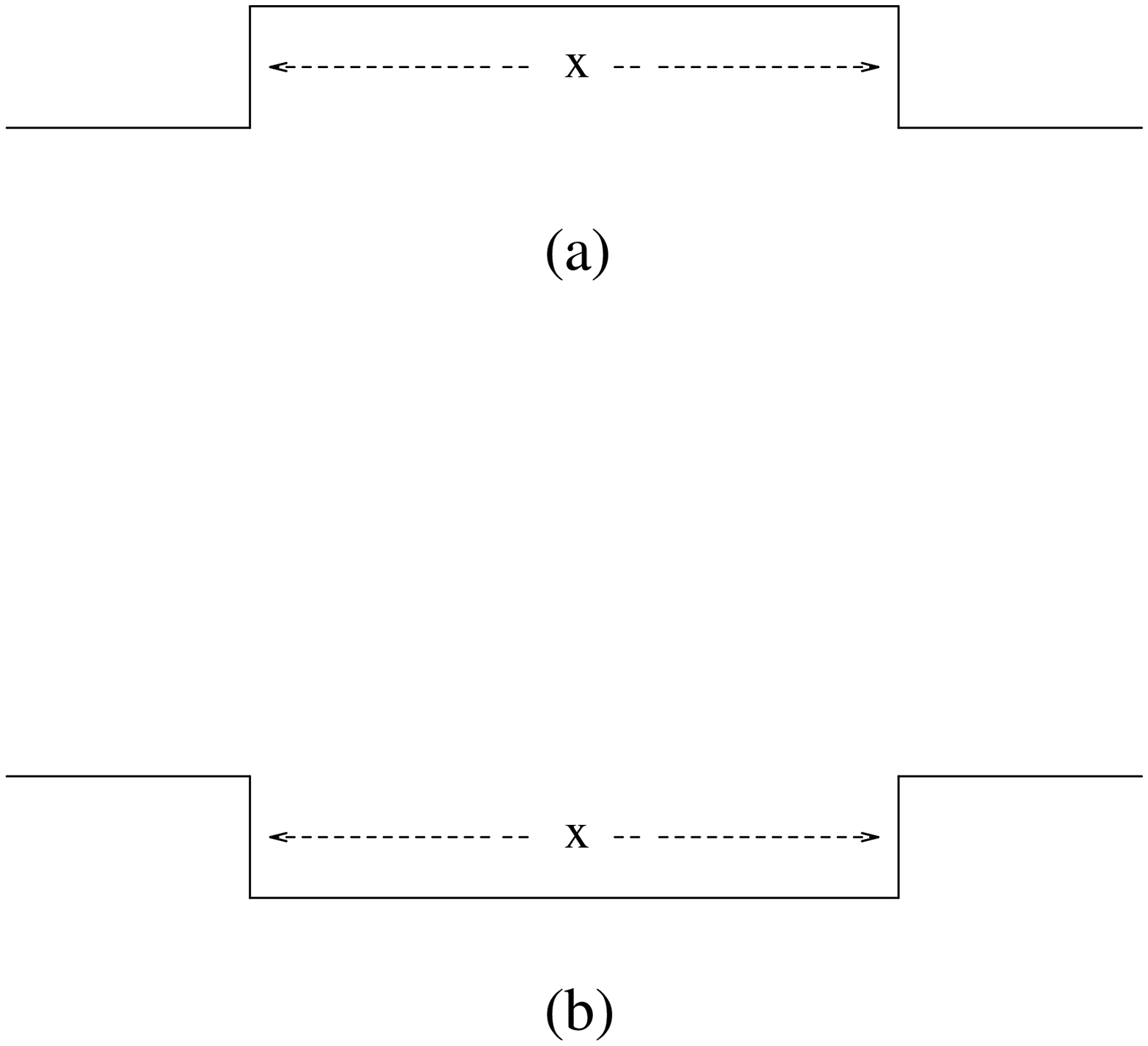}}
\vspace{3cm}
\begin{center}
{\bf Figure 9}
\end{center}
\end{figure}

\newpage
\begin{figure}
\centerline{
\psfig{figure=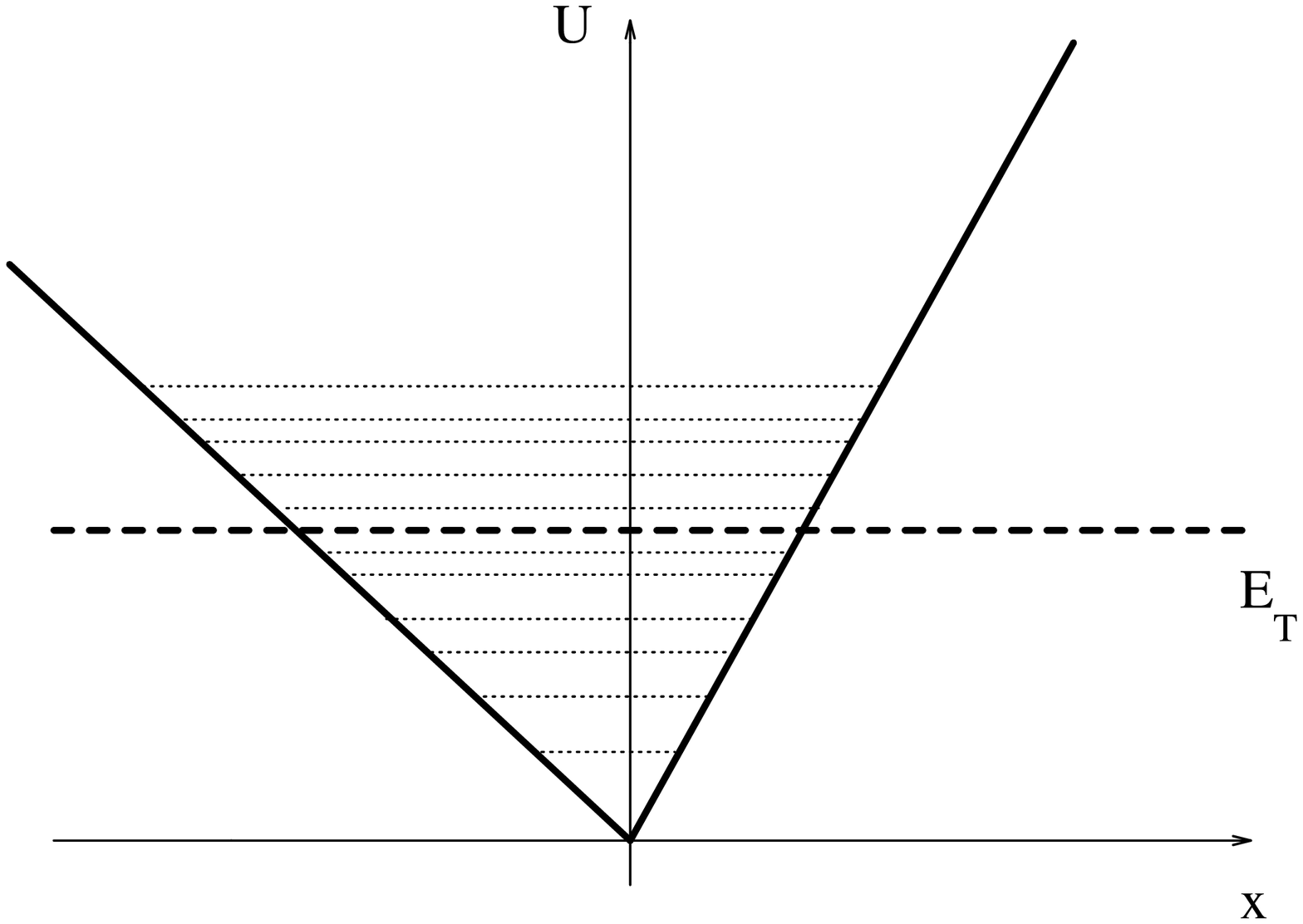}}
\vspace{1cm}
\begin{center}
{\bf Figure 10}
\end{center}
\end{figure}

\end{document}